\documentclass[letterpaper,  amsmath,twocolumn]{revtex4}
\usepackage{amsmath}
\usepackage{graphicx}
\usepackage{dcolumn}
\usepackage{algorithm}
\usepackage{algorithmic}
\begin{document}

\title{Dynamical mean-field theory from a quantum chemical perspective}

\author{Dominika Zgid and Garnet Kin-Lic Chan}
\affiliation{Department of Chemistry and Chemical Biology\\
Cornell University\\Ithaca, NY 14853, USA}
\date{\today}

\begin{abstract}
We investigate the dynamical mean-field theory (DMFT) from a quantum chemical perspective. Dynamical mean-field
theory offers a formalism to extend quantum chemical methods for finite systems
to infinite periodic problems within a local correlation approximation. 
In addition, quantum chemical techniques can be used to construct
 new \textit{ab-initio} Hamiltonians and impurity solvers for DMFT. Here we explore some
 ways in which these things may be achieved. First, we present an informal overview of dynamical mean-field theory to
connect to quantum chemical language. Next
we describe an implementation of dynamical mean-field theory where we start from an \textit{ab-initio} Hartree-Fock Hamiltonian that
avoids  double counting issues present in many applications of DMFT. 
We then explore the use of the  configuration interaction hierarchy in  DMFT as
an approximate solver for the impurity problem. We also investigate some numerical issues of convergence within  DMFT. Our studies
are carried out in the context of the cubic hydrogen model, a simple but challenging test for correlation methods. Finally
we finish with some conclusions for future directions.
\end{abstract}

\maketitle
\section{Introduction}

In molecular quantum chemistry, the use of systematic hierarchies of electron correlation methods to
obtain convergent solutions of the many-electron Schr\"odinger equation has proven very successful. 
For example, the hierarchy of second-order Moller-Plesset perturbation theory (MP2), coupled cluster singles doubles theory (CCSD),
and coupled cluster singles doubles theory with perturbative triples (CCSD(T)) can be used (when strong correlation effects
are absent) to obtain  properties of many small molecules with chemical accuracy \cite{olsen_purplebook}. The computational scalings
of the above methods are respectively $n^5$, $n^6$, and $n^7$, where $n$ is the size of the basis,
 which  seems to limit them to very small systems. However,  local correlation techniques can further be used to 
reduce the above scalings in large systems to $n$, and this has extended the applicability
of such quantum chemical hierarchies to systems  with as many as a thousand atoms \cite{werner_localreview, scuseria_local_correlation, juns_arxiv_paper,ochsenfeld_jcp_2009}. 

Less progress has been made, however, in the use of such quantum chemical hierarchies in
infinite systems such as crystalline solids. We recall briefly the reasons why.
Consider a molecular crystal, where the  molecular unit cell is represented by a basis of $n$ orbitals. 
Assuming $V$ cells in the Brillouin zone of the crystal, the solid is then represented  
by a basis of $nV$ orbitals. 
In   density functional theory (computationally a single-electron theory) 
the cost of the calculation scales as the third power  of the number of orbitals. However,
 translational symmetry  means that one-electron operators (such as the Kohn-Sham Hamiltonian) separate
into $V$ blocks along the diagonal, and the crystal calculation can  be performed for only $V$ times the 
cost of the molecular calculation, rather than $V^3$ times, if translational symmetry were absent.
In correlated calculations, translational symmetry yields a less dramatic advantage. 
For example, for second-order Moller-Plesset perturbation theory, while the molecular calculation scales as $n^5$, the
  scaling of the crystal calculation with translational symmetry is $n^5V^3$,  and  there is still a 
very steep and prohibitive cost dependence on the size of the Brillouin zone \cite{hirata_pccp_2009}. 

Locality of correlation suggests that a formal high scaling with Brillouin zone size can be avoided in physical systems.
(Indeed there are many current efforts underway to explore local correlation methods in the crystal setting) \cite{crystal_mp2, scuseria_periodic_mp2}.
We can then imagine starting with a different picture of a crystal which is more local in nature. Consider  a unit cell in a crystal.
It is  \textit{embedded in a medium}, namely,  the rest of the crystal.  
Translational symmetry implies that the medium consists of the same unit cells as the embedded cell, and thus an appropriate
embedding theory for a crystal should take on a self-consistent nature. If we were to carry out the embedding exactly, we should
not expect any less cost than the full crystal calculation. However, if we  make the assumption that we will neglect (in some manner) 
inter-cell correlations due to locality, then we can expect the high scaling with Brillouin zone size  to vanish, since the theory  takes
on the form of a self-consistent theory for a single unit cell.

Recently,  dynamical mean-field theory (DMFT) has been applied with  success to strongly correlated crystal problems, which
are typically not well described by density functional theory or low-order Green's function techniques \cite{PhysRevB.45.6479,PhysRevLett.69.168,PhysRevLett.62.324,RevModPhys.68.13, antoine_notes, RevModPhys.78.865,Held_review_2007,Held_review_2006}. Note that in this paper, we
will use the term DMFT in a general sense, to mean not only the
single-site variant but also its cluster and multi-orbital extensions~\cite{RevModPhys.77.1027}.
 From one perspective, 
 dynamical mean-field theory can be viewed as a framework which 
realises the self-consistent embedding with local correlation view of a crystal described
 above. DMFT is formulated in the language of Green's functions, and has the form
of a self-consistent theory for the Green's function of a unit cell (which may be a primitive
cell, or more generally a computational supercell). The local correlation approximation
 is   expressed by assuming that the self-energy is local i.e. inter-cell elements of the self-energy vanish, or
in momentum space, that the self-energy is momentum independent. It is important to note that although
correlation effects are neglected between unit-cells, one-electron delocalisation effects between  unit cells are included.
This, together with the self-consistent nature of the embedding distinguishes the physics contained in DMFT
from that in simpler quantum chemical embedding formalisms, such as QM/MM theory \cite{qm_mm_review_friesner}. DMFT has some connections in spirit also to density 
functional embedding methods \cite{wesolowski_warshel, carter_huang_jcp}, although the use of Green's functions avoids 
the need to approximate a non-explicit non-additive kinetic energy functional.

There are several ways in which  DMFT can benefit the  traditional quantum chemical
correlation hierarchy and vice versa. First, DMFT provides a framework through which  quantum chemical
methods for finite systems can be translated to the infinite crystal through
the local correlation approximation, avoiding  the cost of correlated Brillouin zone sampling. (This is true
even for non-size-extensive methods such as configuration interaction, as one is treating
the correlation only within a unit cell and a bath, not the whole crystal simultaneously). The natural way to combine
quantum chemical wavefunction  methods with DMFT is through the discrete bath formulation of DMFT, where 
we need  to determine  the Green's function of a unit cell coupled to a finite non-interacting bath, a so-called impurity problem.
Second, quantum chemistry provides systematic ways to treat  many-body correlations in the DMFT framework. These
quantum chemical solvers are of a different nature to many of the currently used DMFT approximations. 
Finally, quantum chemical methods and basis sets allow us to define the \textit{ab-initio} Hamiltonian and matrix elements  needed
to carry out DMFT calculations in  real systems, while avoiding the empirical parametrisation and double counting corrections that are
currently part of the DFT-DMFT framework.

The current work can be viewed as taking first steps along some of the lines described above. 
We aim to do several things in this paper. First, we  provide an informal description of DMFT 
from an embedding perspective. While we do not introduce new  ideas in this context,
we hope this description may be helpful in forming connections to quantum chemical approximations.
Second, we explore quantum chemical wavefunction correlation methods (more specifically, the configuration interaction hierarchy)
 in the DMFT framework within the discrete bath formulation. These wavefunction methods are used as approximate solvers
for the DMFT impurity problem. Third, we define the DMFT Hamiltonian starting from \textit{ab-initio} Hartree-Fock theory for the crystal, avoiding any double counting or empirical approximations.(Here we point out Ref.~\cite{millis_preprint} the preprint of which appeared as this work
was prepared for submission, which also starts from HF theory
to avoid double counting, though in the different context of DMFT as
applied to a finite system).
Fourth, we explore some of the basic numerics of the DMFT framework, such as the fitting and convergence of the finite
bath approximation, and the convergence of the self-consistency. We explore all these questions in the context of a simple
model system, cubic hydrogen crystal. While a simple system, the correlation in cubic hydrogen can  be tuned from the weak to strong limit as
a function of the lattice spacing, and at least in certain regimes, contains correlation features (such as the three peak
structure of the density states in the intermediate regime) that to date can only be captured within the DMFT framework. 

The structure of the paper is as follows. We begin in section \ref{sec:dmft_overview} with an overview
of the DMFT formalism, starting with a recap of relevant theory of Green's functions, then proceeding
to a general discussion of DMFT self-consistency and embedding, the formulation of the impurity problem and
the  many-body solver, and the definition of the DMFT Hamiltonian starting from Hartree-Fock theory to avoid double counting.
Section \ref{sec:dmft_algorithm} summarises our  implementation of the DMFT algorithm. Section \ref{sec:dmft_results}
describes our exploration of several aspects of the marriage of DMFT and quantum chemistry methods and DMFT numerics
in the cubic hydrogen system, including the use of the configuration interaction hierarchy as a solver, the convergence of the DMFT self-consistency, and the convergence of the DMFT calculations as a function of the bath size.
 We present our conclusions in section \ref{sec:conclusions}.

\section{An informal overview  of  DMFT}

\label{sec:dmft_overview}

\subsection{Summary of Green's function formalism}

\label{sec:summary} 
To keep our discussion self-contained and to establish notation, we begin by recalling some of the basic results from the theory of Green's functions. More detailed exposition of Green's functions can be found, for example, in \cite{Fetter_Walecka}. Given a Hamiltonian $H$ and chemical potential $\mu$, at zero-temperature the Green's function $\mathbf{G}(\omega)$ is defined  as
\begin{align}
G_{ij}(\omega)&=
\langle \Psi_{0} | a_{i} \frac{1}{\omega + \mu - (H-E_{0}) + i0}  a^{\dag}_{j} | \Psi_{0} \rangle  \nonumber \\
&+ \langle \Psi_{0} | a^{\dag}_{j} \frac{1}{\omega +\mu + (H-E_{0}) - i0}  a_{i} | \Psi_{0}\rangle 
  \label{eq:greens_def}
\end{align}
where $i,j$ label the orthogonal one-particle basis, and $\Psi_{0}$ and $E_{0}$ are the ground-state eigenfunction and 
eigenvalue of $H$, respectively.
 $\mathbf{G}(\omega)$  explicitly determines many of the interesting properties
of the system. For example the single-particle density matrix $\mathbf{P}$, electronic energy $E$, and spectral function (density of states) $\mathbf{A}(\omega)$ are given respectively by
\begin{align}
\mathbf{P}&=-i \int_{-\infty}^{\infty} e^{i\omega0_+} \mathbf{G}(\omega) d\omega\label {eq:dm_formula}\\
E&=-\frac{1}{2}i \int_{-\infty}^{\infty} e^{i\omega0_+} \mathrm{Tr} [(\mathbf{h} + \omega ) \mathbf{G}(\omega)] d\omega \label{eq:e_formula}\\
\mathbf{A}(\omega)&= -\frac{1}{\pi} \Im \mathbf{G}(\omega + i0_+) \label{eq:spec_formula}
\end{align}
In general, $\omega$ is  a complex variable. Real $\omega$
corresponds to physical frequencies, and for example, the density of states (\ref{eq:spec_formula}) is defined on the real axis.
However, it is often more convenient to work away from the real axis. For example,
expectation values such as Eqs. (\ref{eq:dm_formula}), (\ref{eq:e_formula}), should  be evaluated on contours away from the real axis to
 avoid singularities in the numerical integration.

In a  crystal, we  assume a localized orthogonal one-particle basis of dimension $n$ in each unit cell. Using translational invariance,
it is sufficient to write the Green's function as
 $\mathbf{G}(\mathbf{R},\omega)$, where $\mathbf{R}$ is the translation vector between unit cells
and for each $\mathbf{R}, \omega$, $\mathbf{G}(\mathbf{R},\omega)$ is an $n \times n$ matrix. 
We shall often refer to the Green's function of a  unit cell in this work as the \textit{local} Green's function.
The local Green's function is then the block of $\mathbf{G}(\mathbf{R},\omega)$ at the origin $\mathbf{R}=0$
and we denote this by $\mathbf{G}(\mathbf{R}_0,\omega)$. The local Green's function determines the local observables,
such as the density matrix of the unit cell, or the local density of states, via formulae analogous to Eqs. (\ref{eq:dm_formula}), (\ref{eq:spec_formula}).
With periodicity, we can also work in the reciprocal  $k$-space. The
$k$-space Green's function $\mathbf{G}(\mathbf{k}, \omega)$ is defined from the Fourier transform
\begin{align}
\mathbf{G}(\mathbf{k},\omega) = \sum_{\mathbf{R}} \mathbf{G}(\mathbf{R},\omega) \exp (i \mathbf{k}\cdot \mathbf{R})
\end{align}
and the local Green's function is obtained from the inverse  transform as
\begin{align}
\mathbf{G}(\mathbf{R}_0,\omega) = \frac{1}{V} \sum_{\mathbf{k}} \mathbf{G}(\mathbf{k}, \omega) \label{eq:dmft_local}
\end{align}
where $V$ is the volume of the Brillouin zone.

When the finite system Hamiltonian is of single particle form, $h=\sum_{ij} h_{ij} a^\dag_i a_j$, the corresponding
non-interacting Green's function is  obtained  from the one-electron matrix $\mathbf{h}$ as
\begin{align}
\mathbf{g}(\omega)=[(\omega + \mu + i0_\pm)\mathbf{1}-\mathbf{h}]^{-1} \label{eq:one_electron_greens}
\end{align}
where we use the convention of lower case  $\mathbf{g}(\omega)$ and $\mathbf{h}(\omega)$ to denote
quantities associated with a non-interacting problem, and the infinitesimal broadening $0_\pm$ is positive
or negative depending on the sign of $\omega$.
In a periodic crystal, we
obtain the non-interacting Green's function in $k$-space from the $k$-space Hamiltonian $\mathbf{h}(\mathbf{k})$
for each $k$ point, 
\begin{align}
\mathbf{g}(\mathbf{k},\omega) = [{(\omega + \mu + i0_\pm)\mathbf{1}-\mathbf{h}(\mathbf{k})}]^{-1} \label{eq:one_electron_greens_k_ortho}
\end{align}

Green's functions $\mathbf{G}(\omega), \mathbf{G}^\prime(\omega)$ corresponding to   different Hamiltonians $\mathbf{H},\mathbf{H}^\prime$ are related 
through frequency dependent one-particle potentials termed  self-energies.
The self-energy $\mathbf{\Sigma}(\omega)$  is  defined via the  Dyson equation as
\begin{align}
\mathbf{\Sigma}(\omega) = {\mathbf{G}^\prime}^{-1}(\omega)- \mathbf{G}^{-1}(\omega) \label{eq:self_energy}
\end{align}
It contains  all the physical effects associated with the perturbation  $\mathbf{H}^\prime-\mathbf{H}$. For example,
we can exactly relate the non-interacting Green's function $\mathbf{g}(\omega)$ from Eq.~(\ref{eq:one_electron_greens}) 
associated with non-interacting Hamiltonian $\mathbf{h}$,
and the  interacting Green's function $\mathbf{G}(\omega)$  associated with interacting Hamiltonian $\mathbf{H}$,
 through a Coulombic self-energy.
From the explicit form of the non-interacting Green's function $\mathbf{g}(\omega)$, the Dyson equation in this case is
\begin{align}
\mathbf{G}^{-1}(\omega) = (\omega+\mu + i0_\pm)\mathbf{1} - \mathbf{h} - \mathbf{\Sigma}(\omega) \label{eq:green_no_k_dept_anywhere_ortho}
\end{align}
In a periodic system,  the above equation holds at each $k$
where  the self-energy $\mathbf{\Sigma}(\mathbf{k},\omega)$ now also acquires a $k$-dependence, 
\begin{align}
\mathbf{G}^{-1}(\mathbf{k}, \omega) = (\omega+\mu +i0_\pm)\mathbf{1} - \mathbf{h}(\mathbf{k}) - \mathbf{\Sigma}(\mathbf{k},\omega) \label{eq:green_in_k_space},
\end{align}
and  the local Green's function becomes
\begin{align}
\mathbf{G}(\mathbf{R}_{0}, \omega)=\frac{1}{V} \sum_{\mathbf{k}} [{(\omega+\mu +i0_\pm)\mathbf{1} - \mathbf{h}(\mathbf{k}) - \mathbf{\Sigma}(\mathbf{k},\omega) }]^{-1} \label{eq:green_in_real_space_k_se_ortho}.
\end{align}
In general, it is convenient to relax the assumption of orthogonality of the one-particle basis, for example, to work with an
 atomic orbital basis. For this, the unit matrix  $\mathbf{1}$ in the above formulae should
be replaced by a general overlap matrix $\mathbf{S}$, e.g. Eq. ~(\ref{eq:green_in_real_space_k_se_ortho})
becomes
\begin{align}
\mathbf{G}(\mathbf{R}_{0}, \omega)=\frac{1}{V} \sum_{\mathbf{k}} [{(\omega+\mu +i0_\pm)\mathbf{S(k)} - \mathbf{h}(\mathbf{k}) - \mathbf{\Sigma}(\mathbf{k},\omega) }]^{-1} \label{eq:green_in_real_space_k_se},
\end{align}
In addition expectation values must be suitably modified. For example, the local spectral function $\mathbf{A}(\mathbf{R}_0, \omega)$
is given by
\begin{align}
\mathbf{A}(\mathbf{R}_0, \omega) = \frac{1}{V} \Im \sum_{\mathbf{k}} \mathbf{G}(\mathbf{k}, \omega+i0_+) \mathbf{S}(\mathbf{k})\label{eq:overlap_spectral}
\end{align}
As our calculations in this work use a non-orthogonal basis, we will henceforth use expressions 
with  explicit overlap dependence.

\subsection{DMFT equations}

In DMFT, the central quantity is the local Green's function $\mathbf{G}(\mathbf{R}_0,\omega)$ (the Green's function
of the unit cell) which is determined in a self-consistent way, including the embedding effects of the crystal
within a local self-energy (correlation) assumption. 
Here we describe  how the DMFT framework and the local self-energy assumption
and self-consistency are established. Of course, we recommend  that the reader also consult one of the many excellent review articles for
further discussion and illumination of the DMFT formalism \cite{RevModPhys.68.13, antoine_notes, RevModPhys.78.865,Held_review_2007,Held_review_2006}.

From Eq.~(\ref{eq:green_in_real_space_k_se}), we observe that $\mathbf{G}(\mathbf{R}_0, \omega)$  can be calculated 
 if we have the exact Coulomb self-energy $\mathbf{\Sigma}(\mathbf{k}, \omega)$. However, determining $\mathbf{\Sigma}(\mathbf{k},
\omega)$ requires solving the many-body problem for the whole crystal.
Thus the idea in DMFT is to approximate $\mathbf{\Sigma}(\mathbf{k}, \omega)$ 
by one of its main components, the local self-energy  $\mathbf{\Sigma}(\omega)$, in essence, a local correlation approximation.
Formally, this is the contribution 
to the self-energy of  skeleton diagrams in the Green's function perturbation theory
where the Coulomb interaction has all local indices, i.e. all indices local to a single unit cell.
The DMFT approximation  neglects the $k$-dependence of the self-energy.
 In real-space, this corresponds to
 neglecting  off-diagonal terms of the self-energy between unit cells. The local approximation
is plausible due to the local nature of correlation, and in fact
as the physical dimension or local coordination number $D \to \infty$, the  approximation becomes exact \cite{RevModPhys.68.13}.
With the DMFT local  approximation, the local Green's function defined in Eq.~(\ref{eq:dmft_local}) is simply
\begin{align}
\mathbf{G}(\mathbf{R}_0,\omega)=\frac{1}{V}\sum_{\mathbf{k}}[{(\omega+\mu+i0_\pm)\mathbf{S}(\mathbf{k}) - \mathbf{h}(\mathbf{k}) - \mathbf{\Sigma}(\omega)}]^{-1} \label{eq:dmft_local_greens}
\end{align}

Now $\mathbf{\Sigma}(\omega)$ is formally defined by
 contributions of only the local Coulomb interaction to the local Green's function. However,  this is still a 
many-body problem. In DMFT, we usually reformulate the determination of $\mathbf{\Sigma}(\omega)$ in terms of the many-body solution 
of an embedded, or \textit{impurity}, problem where we view the unit cell as an impurity embedded
in a bath of the surrounding crystal. (The impurity nomenclature originates from impurity problems in condensed matter such
as the Kondo and Anderson models,
which informed some of the early work in DMFT). Within this impurity mapping, the many-body determination of the 
Green's function of the embedded unit cell or impurity 
Green's function $\mathbf{G}_{imp}(\omega)$, defines the local self-energy $\mathbf{\Sigma}(\omega)$.

 We  discuss the impurity problem, and impurity solvers to obtain the self-energy, in  more detail
the next section. We focus for now on how the self-consistent embedding is established in DMFT.
For the theory to be consistent, the impurity Green's function
(i.e. the Green's function of the embedded unit cell in the impurity model)
should be  equivalent to the actual local Green's function of the crystal, at least within the local self-energy approximation. This means
at self-consistency, 
\begin{align}
\mathbf{G}_{imp}(\omega)=\mathbf{G}(\mathbf{R}_0,\omega) \label{eq:scf}
\end{align}
The embedding to achieve the equality (\ref{eq:scf}) can be enforced 
through an embedding self-energy, the hybridization $\mathbf{\Delta}(\omega)$. 
The Dyson equation relating the impurity Green's function and the self-energy and hybridization is then
\begin{align}
\mathbf{G}_{imp}(\omega)^{-1} = (\omega+\mu+i0_\pm)\mathbf{S} - \mathbf{h}_{imp} - \mathbf{\Sigma}(\omega) - \mathbf{\Delta}(\omega) \label{eq:gimp}
\end{align}
where $h_{imp}$ is a one-electron Hamiltonian in the unit cell. Once we have solved the many-body impurity problem
to obtain $\mathbf{G}_{imp}$, Eq. (\ref{eq:gimp})  defines the local self-energy through
\begin{align}
\mathbf{\Sigma}(\omega) = (\omega+\mu+i0_\pm)\mathbf{S} - \mathbf{h}_{imp} -\mathbf{\Delta}(\omega) - \mathbf{G}_{imp}(\omega)^{-1} \label{eq:self_energy_defn}
\end{align}
The hybridization $\mathbf{\Delta}(\omega)$ can also be defined through a similar equation from the local Green's function, obtained
 from Eq.~(\ref{eq:dmft_local_greens}) 
\begin{align}
\mathbf{\Delta}(\omega) = (\omega+\mu+i0_\pm)\mathbf{S} - \mathbf{h}_{imp} -\mathbf{\Sigma}(\omega) - \mathbf{G}(\mathbf{R}_0,\omega)^{-1} \label{eq:hybrid_defn}
\end{align}

Schematically therefore, for a given hybridization $\mathbf{\Delta}(\omega)$, solution of the  impurity problem 
yields $\mathbf{G}_{imp}(\omega)$ and the local self-energy $\mathbf{\Sigma}(\omega)$ 
\begin{align}
\mathbf{\Delta}(\omega) \stackrel{\text{impurity solver}}{\to} \mathbf{G}_{imp}(\omega) \to \mathbf{\Sigma}(\omega) \label{eq:self_e_scheme}
\end{align}
while given the local self-energy, Eq. (\ref{eq:dmft_local_greens}) yields
the local Green's function and the hybridization
\begin{align}
 \mathbf{\Sigma}(\omega) \to \mathbf{G}(\mathbf{k},\omega) \to \mathbf{G}(\mathbf{R}_0,\omega) \to \mathbf{\Delta}(\omega) \label{eq:hybrid_scheme}
\end{align}
Eq.~(\ref{eq:hybrid_scheme}) and Eq. (\ref{eq:self_e_scheme}) thus form a self-consistent pair of equations
for the self-energy and hybridization that should be
iterated to convergence. These are the DMFT self-consistent equations. At the solution point, the impurity Green's function and local Green's function, are identical as in Eq. (\ref{eq:scf}).

We note here that the Green's functions $\mathbf{G}(\mathbf{R}_0, \omega), \mathbf{G}_{imp}(\omega)$, and
the self-energy and hybridisation $\mathbf\Sigma(\omega), \mathbf{\Delta}(\omega)$ are smooth functions away from the real axis. For this
reason,  the impurity problem and the numerical implementation of self-consistency are always considered on the imaginary axis rather than the real axis. Once
the self-consistency Eq. (\ref{eq:scf}) has been reached on the imaginary axis, analyticity guarantees equivalence of the Green's functions in the whole
complex plane. One can then use the converged $\mathbf{\Delta}(\omega)$ (continued to the real axis) to recalculate properties
along the real axis, such as spectral functions, as needed. (Many quantities, such as density matrices, require
only information along the imaginary axis, however).

We recap the main physical effects contained within  the DMFT treatment - 
local Coulomb interaction effects are included in each unit cell and replicated
throughout the crystal, in a self-consistent way which takes into account the embedding of each unit cell
in an environment of the others. Long-range Coulomb terms are not included in the theory although they can be systematically
added. In section \ref{sec:hamiltonian} we describe how the long-range terms can be treated at the mean-field level. 

Note that we have assumed in the above that we are working at a fixed $\mu$. Normally, however, 
we are interested not in fixed $\mu$, but in some fixed particle
number of the crystal per unit cell, $N_0(\mathbf{R}_0)$. As $\mathbf{\Sigma}(\omega)$ changes,
$N(\mathbf{R}_0)$, the current particle number in the crystal unit cell, given by (using Eqs. (\ref{eq:dm_formula}) and (\ref{eq:dmft_local_greens}))
\begin{widetext}
\begin{align}
N(\mathbf{R}_0) = -\frac{i}{V}\int_{-\infty}^{\infty} e^{i\omega0_+} \mathrm{Tr} \left[\sum_k \mathbf{S}(\mathbf{k}) [(\omega+\mu+i0_\pm)\mathbf{S}(\mathbf{k}) - \mathbf{h}(\mathbf{k}) - \mathbf{\Sigma}(\omega)]^{-1} \right]d\omega \label{eq:particle_number}
\end{align}
\end{widetext}
will change. Thus together with the self-consistency, the chemical potential $\mu$ must be adjusted such that
$N(\mathbf{R}_0)=N_0(\mathbf{R}_0)$. The full DMFT algorithm to do so is summarised in section \ref{sec:dmft_algorithm}.

We now turn to consider the many-body impurity problem and methods for its solution.

\subsection{The impurity problem and solver in the discrete bath formulation}
\label{sec:solver}
The purpose of the impurity formulation is to obtain an impurity Green's function $\mathbf{G}_{imp}(\omega)$ 
and a corresponding self-energy $\mathbf{\Sigma}(\omega)$ that
describes the effects of the local Coulomb interaction in the presence of the hybridization $\mathbf{\Delta}(\omega)$.
In general, due to its many-body nature, the impurity problem cannot be solved exactly. The approximate method used to solve the impurity problem is known as the impurity solver. 

There are two  formulations in which an impurity solver can work \cite{RevModPhys.68.13}. In the first one
the impurity Green's function is expressed as  a functional integral, and its determination
is a problem of high-dimensional integration. This is typically performed using Monte Carlo methods such 
as  Hirsch-Fye~\cite{Hirsh_Fye_1986} or continuous time quantum Monte Carlo methods~\cite{Rubtsov_Lichtenstein_weak_coupling_2004,
Rubtsov_Lichtenstein_weak_coupling_2005,Werner_hybrid_expansion_2006,Gull_ctaux_2008}. In this formulation, the bath is infinite and  one does not deal with it explicitly since it can be integrated out thus avoiding any bath discretization error. 
These methods are  powerful but suffer in general from a sign problem, as well as difficulties in obtaining
quantities on the real frequency axis (such as the spectral function) which requires analytic continuation.
We will not discuss the  Monte Carlo formulations of the solver
further here, but we refer the reader to an excellent review~\cite{Gull_review_2010}.

The second formulation describes an impurity model with an explicit finite, discrete bath. Here the  idea is to view the 
 hybridization $\mathbf{\Delta}(\omega)$  as arising from a one-electron coupling between
the impurity orbitals (orbitals of  the unit cell)
 and a fictitious  finite non-interacting bath. The relevance of the formulation with discrete bath here is that
the  determination of the impurity Green's function reduces to the determination of the Green's function
of a finite problem, and this can be tackled using standard quantum chemistry wavefunction techniques which
avoid the sign problem
encountered in Monte Carlo based solvers. We can
view then such an discrete bath formulation as providing a way to extend quantum chemical methods for finite systems
to treat the infinite crystal, within the DMFT approximation of a local self-energy.

Denoting the local orbitals by $i, j, \ldots$, 
and  bath orbitals by $p, q \ldots$, we can write an impurity Hamiltonian for the impurity orbitals and the
fictitious non-interacting bath  as
\begin{widetext}
\begin{align}
H_{imp+bath} = \sum_{ij} t_{ij} a^\dag_i a_j + \frac{1}{2} \sum_{ijkl} w_{ijkl} a^\dag_i a^\dag_j a_l a_k
+ \sum_{ip} V_{ip} (a^\dag_i a_{p} + a^\dag_{p} a_i) + \sum_{p} \epsilon_{p}  a^\dag_{p}  a_{p} \label{eq:impurity_model_h}
\end{align}
\end{widetext}
The non-interacting bath yields a hybridization $\mathbf{\Delta}(\omega)$ for the impurity orbitals of the form
\begin{align}
\Delta_{ij}(\omega)&=\sum_{p} \frac{V_{ip}^* V_{jp}}{\omega - \epsilon_{p}} \label{eq:bath_representation}
\end{align}
In general, we assume that  physical $\mathbf{\Delta}(\omega)$ can be approximately represented
 in terms of the non-interacting bath by fitting the couplings $V$ and the energies $\epsilon$, and this is generally found to be true.
This resembles the assumption of non-interacting $v$-representability of the density in density functional theory.
Fortunately, the convergence of (\ref{eq:bath_representation}) with respect to the number of bath orbitals is quite rapid; one does not need
a bath the size of the entire crystal to obtain a good representation of the hybridization. (Recall that the fit to the bath
is always carried out on the imaginary frequency axis, where $\mathbf{\Delta}(\omega)$ is very smooth).

The form of the bath hybridisation in Eq. (\ref{eq:bath_representation}) requires that $\lim_{\omega \to \infty} \mathbf{\Delta}(\omega) \to 0$.
While this is true of physical hybridisations  in an orthogonal basis, the case of a non-orthogonal basis requires a little more care, as discussed for example, in Ref.~\cite{RevModPhys.78.865}. Rearranging  Eq. (\ref{eq:gimp})
and inserting the definition of the local Green's function, we see that the hybridisation is given by
\begin{align}
\mathbf{\Delta}(\omega)=(\omega+\mu+i0_\pm)\mathbf{S}_{imp} - \mathbf{h}_{imp}  - \mathbf{\Sigma}(\omega) -  \mathbf{G}(\mathbf{R}_0,\omega)^{-1}
\end{align}
The definition of the impurity overlap matrix $\mathbf{S}_{imp}$ and impurity one-electron Hamiltonian $\mathbf{h}_{imp}$ can be viewed as adjustable
as the equality of the impurity Greens function and local crystal Green's function,
 can be maintained through appropriate definitions of the
hybridisation and self-energy in Eq. (\ref{eq:gimp}). Consequently, we  choose $\mathbf{S}_{imp}$ and $\mathbf{h}_{imp}$ to ensure that the hybridisation
can be represented by the form Eq. (\ref{eq:bath_representation}). Expanding the denominator in powers of $1/\omega$, we find
that to ensure $\mathbf{\Delta}(\omega)$ vanishes like $1/\omega$, we should define the impurity overlap and one-electron Hamiltonian as
\cite{RevModPhys.78.865}
\begin{align}
\mathbf{S}_{imp}&= \left[\frac{1}{V}\sum_{\mathbf{k}}\mathbf{S}^{-1}(\mathbf{k})\right]^{-1}, \label{eq:savrasov_overlap} \\
\mathbf{h}_{imp}&=\mathbf{S}_{imp}\left[ \sum_{\mathbf{k}}\mathbf{S}^{-1}(\mathbf{k})[\mathbf{h(k)}+\mathbf{\Sigma}_{\infty}]\mathbf{S}^{-1}(\mathbf{k})\right]\mathbf{S}_{imp}-\mathbf{\Sigma}_{\infty} \label{eq:savrasov_hamiltonian},
\end{align}
where $\mathbf{\Sigma}_{\infty}=\mathbf{\Sigma}(\infty)$.

Now that  we have defined a finite Hamiltonian for the  impurity and a finite bath,
the determination of the impurity Green's function $\mathbf{G}_{imp}(\omega)$
is   the determination of the Green's function of a finite problem.
$\mathbf{G}_{imp}(\omega)$ is defined through
 Eq. (\ref{eq:greens_def}) with the impurity Hamiltonian,
\begin{align}
&G_{ij}(\omega)= \nonumber \\
& \langle \Psi_{0} | a_{i} \frac{1}{\omega + \mu - (H_{imp+bath}-E_{imp+bath}) + i0}  a^{\dag}_{j} | \Psi_{0} \rangle  \nonumber \\
&+ \langle \Psi_{0} | a^{\dag}_{j} \frac{1}{\omega +\mu + (H_{imp+bath}-E_{imp+bath}) - i0}  a_{i} | \Psi_{0}\rangle 
  \label{eq:greens_def2}
\end{align}
where $i,j$ denote the impurity orbitals, i.e. the local orbitals of the unit cell, and $E_{imp+bath}, \Psi_0$ are the ground-state eigenvalue
and eigenfunction of $H_{imp+bath}$.
Both $\Psi_0$ and the corresponding $\mathbf{G}_{imp}(\omega)$ can be determined through   wavefunction techniques familiar in quantum chemistry.

One subtlety
is that the finite problem $\Psi_0$ is determined for some fixed particle number $N_{imp+bath}$ (and spin, say). In principle, 
at zero temperature, we should use the $N_{imp+bath}^{min}$ (and spin) which minimises  $E_{imp+bath}$ for the given chemical potential $\mu$. This means
that we have to carry out a search over these quantum numbers. Of course $\mu$ and $\mathbf{\Delta}(\omega)$ are
also changing in the DMFT iterations, and thus in the discrete bath formulation, the impurity model is a function
of  $N_{imp+bath}$ (and other quantum numbers), $\mu$, and $\mathbf{\Delta}(\omega)$.
 The structure of the full  self-consistency involving these variables is summarised in the DMFT algorithm in section \ref{sec:dmft_algorithm}.

A popular approach in existing DMFT applications is to use full configuration 
interaction (FCI) called exact diagonalization (ED) in solid state physics community to solve for $\Psi_0$ and $\mathbf{G}_{imp}(\omega)$ \cite{RevModPhys.68.13,PhysRevLett.72.1545}. From a DMFT perspective, the advantage of this approach compared
to Monte Carlo techniques is that it provides direct access to the calculation of the Green's function on the real axis, and consequently the spectral function,
without the need to perform analytic continuation as is used in Monte Carlo solvers. 
In addition, there is no sign problem. 
However, FCI
is naturally limited to very small numbers of impurity and bath orbitals, and the cost of evaluating
the Green's function (typically at several hundred frequencies) means that such calculations are orders of magnitude
more expensive than typical ground-state FCI calculations for molecules. One way to avoid this limitation is to employ the
various systematic quantum chemistry wavefunction hierarchies as impurity solvers.
We will investigate one such simple approximate solver,  the configuration interaction hierarchy, in section \ref{sec:cisolver}.

\subsection{Eliminating double counting in  DMFT   through Hartree-Fock theory}
\label{sec:hamiltonian}
In current applications of DMFT to real materials, it is common to combine DMFT with a density functional derived Hamiltonian,
the so-called DFT-DMFT approximation \cite{RevModPhys.78.865,Held_review_2007,Held_review_2006}. Within
this formalism, one does
not work  with a strict  \textit{ab-initio}  Hamiltonian, 
but  rather with a model  Hamiltonian 
\begin{align}
H_{imp} = H_{DFT} + \frac{1}{2}\sum_{ijkl \in act} w_{ijkl} a^\dag_i a^\dag_j a_l a_k - H_{d.c.} \label{sec:dft_dmft_h}
\end{align}
where $H_{DFT}$ is the  sum of  one-electron Kohn-Sham operators
and $H_{d.c.}$ is a double-counting correction (see below).
The  two electron interaction $w_{ijkl}$ is chosen to sum  over a set of active orbitals in the 
computational unit cell.  In transition metal applications, 
these are usually a minimal basis of $d$ or $f$ valence 
 orbitals, the idea being that the Coulomb interaction in these orbitals should be treated with the explicitly many-body 
DMFT framework, rather than within a DFT functional. While $w_{ijkl}$ may be  obtained from \textit{ab-initio} Coulomb integrals \cite{some_papers_where_this_is_done_eg_lichtenstein_using_Slater_orbitals, aryasetiawan_gw_dmft} or
derived via e.g. constrained DFT calculations \cite{PhysRevLett.53.2512,constranined_dft2}, they are best regarded in this approach as  semi-empirical  parameters. 
The advantage of using DMFT in only an active space is that  delocalised, itinerant electrons are 
 well treated by existing exchange-correlation functionals, and not well-treated
within the DMFT framework which neglects non-local correlations, 
while  the many-body DMFT framework allows a systematic approach to 
   high order strong correlations in the $d$ and $f$ shells. The adjustment of $w_{ijkl}$ further allows one to account for effective
 screening of the active space Coulomb matrix elements by long-range correlations. The DFT-DMFT approach has been  successful in
 reproducing many properties of strongly correlated materials and an excellent description of the possible applications and the way of dealing with the 
double counting correction can be found in Ref.~\cite{Held_review_2007,Held_review_2006,Karolak201011}. However, there are obvious drawbacks. In particular,
the Hamiltonian  may be considered to be uncontrolled on two levels. Firstly, 
since exchange-correlation effects in DFT are not separated between different orbitals, there is a double
counting of the Coulomb interaction in $H_{DFT}$ and $w$. This is the origin of
the   double-counting correction $H_{d.c.}$ which must be adjusted empirically. The double counting problem is similar to that encountered
in molecular quantum chemistry when DFT is combined with active space wave function methods \cite{savin_ci_dft}. Secondly,  the
use of a parametrized form for $w_{ijkl}$ must also be regarded  as unsystematic.

In the current work,  we take a more quantum chemical approach to DMFT where we try to retain a strict diagrammatic
control over the approximations made. This can be achieved by starting with a Hartree-Fock description of the crystal.
Within each unit cell we identify an active space, typically a set of localized atomic orbitals. (In fact, in the application
to cubic hydrogen in this work, all the orbitals in the unit cell will be active). Then,
we use DMFT to treat the active space Coulomb interaction while the remaining Coulomb interactions (e.g.
long-range Coulomb interactions between unit cells, as well the interactions between the active and inactive orbitals)
are treated through the Hartree-Fock mean-field. The  Hamiltonian in the active space treated within DMFT therefore
takes the form
\begin{align}
H_{imp} = \sum_{ij \in act} (f_{ij} - \tilde{f}_{ij}) a^\dag_i a_j  + \frac{1}{2}\sum_{ijkl \in act} 
w_{ijkl} a^\dag_i a^\dag_j a_k a_l \label{eq:himp_gen}
\end{align}
where the $\tilde{f}_{ij}$ terms represents the exact subtraction of  the  active-space Hartree-Fock density matrix $\mathbf{P}^{HF}$, contribution 
to the mean-field Coulomb treatment
\begin{align}
\tilde{f}_{ij} = \sum_{kl \in act} P^{HF}_{kl} (w_{iklj} - w_{ilkj}) \label{eq:h_subtraction}
\end{align}
This subtraction exactly eliminates any double counting between the mean-field and DMFT treatments. 
Note that while the inactive Coulomb interactions (such as the long-range Coulomb interactions) 
are only treated at the Hartree-Fock level (which is
 a severe approximation in many solids) the mean-field treatment may be viewed as the lowest
level of a hierarchy of perturbation treatments of these interactions, and is thus systematically improvable. 
Ref.~\cite{millis_preprint}, whose preprint appeared as this work was prepared for
submission, also explores a Hartree-Fock starting point to avoid
double counting, but in the
context of DMFT applied to finite systems.

\section{DMFT algorithm} \label{sec:dmft_algorithm}

\begin{center}
\begin{algorithm}
\begin{algorithmic}[1]
\FORALL  {$N_{imp+bath}$}
\WHILE {$N({\bf{R}}_0) \neq N_0({\bf{R}}_0)$}
\STATE   Choose new $\mu$ (e.g. by bisection)
\STATE   Perform DMFT self-consistency for $\mathbf{\Sigma}(\omega)$, $\mathbf{\Delta}(\omega)$. 
\STATE   Calculate $E_{imp+bath}$ 
\STATE   Calculate $N(\mathbf{R}_0)$
\ENDWHILE
\ENDFOR
\STATE Choose $N_{imp+bath}^{min}$ that minimises $E_{imp+bath}$ 
\STATE For $N_{imp+bath}^{min}$ and the corresponding $\mu$, $\mathbf{\Delta}(\omega)$ and impurity model,
calculate $\mathbf{G}(\mathbf{R_0}, \omega)$ including quantities on the real axis e.g. spectral functions
\end{algorithmic}
\caption{General DMFT  loop structure. Note that the DMFT self-consistency is carried out on the imaginary frequency axis.\label{algo:one}}
\end{algorithm}
\end{center}
\begin{algorithm}

\begin{algorithmic}[1]
\STATE  Obtain Hartree-Fock Fock matrix $\mathbf{f}(\mathbf{k})$, overlap matrix $\mathbf{S}(\mathbf{k})$, 
density matrix $\mathbf{P}(\mathbf{R}_0)$, and initial guess for $\mathbf{\Delta}(\omega)$.
\WHILE {$||\mathbf{\Sigma}(\omega)-\mathbf{\Sigma}^{old}(\omega)|| > \tau$}
\STATE {Construct Hamiltonian for impurity orbitals with overlap correction (using $\mathbf{\Sigma}(\infty)$)}
\STATE {Construct bath representation from $\mathbf{\Delta}(\omega)$}
\STATE {Calculate impurity Greens function  and new self-energy $\mathbf{\Sigma}(\omega)$}
\STATE {Update self-energy $\mathbf{\Sigma}(\omega)$, $\mathbf{\Sigma}^{old}(\omega)$}
\STATE {Update $\mathbf{\Delta}(\omega)$}
\ENDWHILE
\end{algorithmic}
\caption{DMFT self-consistency for $\mathbf{\Delta}(\omega), \mathbf{\Sigma}(\omega)$. Note that all calculations are done
on the imaginary frequency axis. \label{algo:two}}
\end{algorithm}

We now summarize the DMFT algorithm  in our current implementation, following  the basic ideas outlined in the earlier sections.
We have implemented our algorithm in a custom code that interfaces to the \textsc{Crystal} Gaussian based periodic code~\cite{crystal} as
well as the \textsc{Dalton}  molecular code \cite{dalton}. 
We recall that within the formulation with discrete bath, the impurity model
is defined as a function 
 of three variables: $N_{imp+bath}$ (particle number of the impurity model), $\mu$ (chemical potential), and the hybridisation $\mathbf{\Delta}(\omega)$
which defines a bath parametrisation. 
All three have to be determined self-consistently together. At the solution point of the DMFT algorithm, $N_{imp+bath}$ minimises
 the ground-state energy of the impurity model $E_{imp+bath}$ (section \ref{sec:solver}), $\mu$ yields
the correct particle number per unit cell of the crystal $N(\mathbf{R_0})$ (Eq. \ref{eq:particle_number}), and $\mathbf{\Delta}(\omega)$ satisfies the DMFT
self-consistency conditions (\ref{eq:self_e_scheme}), (\ref{eq:hybrid_scheme}).
The high-level loop structure of the algorithm is summarised in algorithm \ref{algo:one}.
The individual steps are
\begin{itemize}
\item [1.] Loop over possible particle numbers $N_{imp+bath}$ of the impurity model (to determine
 $N_{imp+bath}$ which minimises the impurity model energy $E_{imp+bath}(N_{imp+bath})$). (In principle
we should  search over spin, but we do not do this is in general in our applications here).
\item [2.] For each $N_{imp+bath}$, search over chemical potential $\mu$ (e.g. by bisection)
to satisfy the crystal unit cell particle number constraint $N(\mathbf{R_0}) = N_0(\mathbf{R_0})$.
\item [3.-6.] For given $\mu, N_{imp+bath}$, carry out the DMFT self-consistent loop to
determine $\mathbf{\Sigma}(\omega), \mathbf{\Delta}(\omega)$ and the impurity ground state energy $E_{imp+bath}$. Note
that all calculations are here done on the imaginary frequency axis.
\item [9.-10.] Determine  $N_{imp+bath}$ which led to the lowest $E_{imp+bath}$. Using
the corresponding $\mu$ and hybridisation parametrisation, which satisfy the crystal particle number
constraint and the DMFT self-energy self-consistency equations, recalculate the local Greens function $G(\mathbf{R}_0, \omega)$
and other desired observables, e.g. the local spectral function $\mathbf{A}(\omega)$ along the real axis.
\end{itemize}

The DMFT self-consistent loop for $\mathbf{\Sigma}(\omega)$, $\mathbf{\Delta}(\omega)$ constitutes 
the core part of the algorithm. It is summarised in algorithm \ref{algo:two}. 
The individual steps are
\begin{itemize}
\item [1.] Initialisation. Perform a Hartree-Fock (HF) calculation 
on the crystal  in a local basis. Extract the converged Fock matrix $\mathbf{f}(\mathbf{k})$ and 
 overlap matrix $\mathbf{S}(\mathbf{k})$ in  $k$-space, 
and the Hartree-Fock unit-cell density matrix $\mathbf{P}(\mathbf{R}_0)$. The $k$-space 
Fock and overlap matrices are then used to construct their real-space analogs in the unit-cell.
\item [2.] Begin DMFT self-consistent loop until convergence
in the self-energy (to within a threshold $\tau$) is reached. 
\item [3.] Impurity Hamiltonian construction. Construct the impurity orbital part of the Hamiltonian.
The two-body integrals $w_{ijkl}$ are computed in the same
local basis as used in the crystal calculation. The one-body Hamiltonian
for the impurity orbitals $\mathbf{h}_{imp}$ is defined as in Eq. (\ref{eq:h_subtraction}) using the
exact subtraction of the mean-field Coulomb treatment
i.e. $\mathbf{h}_{imp} = \mathbf{f}(\mathbf{R}_0) - \tilde{\mathbf{f}}(\mathbf{R}_0)$,
while the overlap of the impurity orbitals is taken as the overlap in the unit-cell, $\mathbf{S}_{imp}=\mathbf{S}(\mathbf{R}_0)$.
Finally, $\mathbf{h}_{imp}$ and $\mathbf{S}_{imp}$ 
are corrected as  in Eqs. (\ref{eq:savrasov_overlap}), (\ref{eq:savrasov_hamiltonian}).
\item [4.] Bath construction. From the hybridisation  $\mathbf{\Delta}(\omega)$,
 obtain the bath Hamiltonian parametrisation by fitting.
In the first iteration, the hybridisation is fitted to
the Hartree-Fock hybridisation, defined as
\begin{align}
\mathbf{\Delta}^{HF}(\omega) =& 
({\omega+\mu+i0_{\pm}}) \mathbf{S}_{imp} - \mathbf{h}_{imp}+ \\
 &-  \left[\frac{1}{V}\sum_k (\omega+\mu+i0_{\pm}) \mathbf{S}(\mathbf{k}) - \mathbf{f}(\mathbf{k})\right]^{-1} \nonumber
\end{align}
This provides a good guess for the DMFT algorithm.
Further details of the bath fitting algorithm
are given in section \ref{sec:fitting} and in the appendix.
\item [5.] Calculate the ground-state wavefunction 
of the impurity problem (for given $N_{imp+bath}$). Then
calculate the impurity Green's function on the imaginary axis 
using a  truncated configuration interaction solver, described
in section \ref{sec:cisolver}.
\item [6.-7.] Update the self-energy $\mathbf{\Sigma}(\omega)$ and hybridisation $\mathbf{\Delta}(\omega)$ defined 
through Eqs. (\ref{eq:self_energy}), (\ref{eq:hybrid_defn}). For better convergence,
the self-energy is updated in a damped fashion, $\mathbf{\Sigma}(\omega) \gets (1-\alpha) \mathbf{\Sigma}(\omega) + \alpha \mathbf{\Sigma}^{old}(\omega)$, where $0< \alpha < 1$.
\end{itemize}

\section{Benchmark DMFT studies}

\label{sec:dmft_results}

We now proceed to our benchmark DMFT studies.
 In particular, we investigate
\begin{enumerate}
\item the preliminary combination of quantum chemical and DMFT ideas, using
 the configuration interaction (CI) hierarchy as a solver for the DMFT impurity problem (or conversely,  using DMFT
 to extend truncated CI variants to treat the infinite crystal), starting from an \textit{ab-initio} Hartree-Fock DMFT Hamiltonian,
\item the numerical behaviour of the DMFT algorithm, including convergence of the self-consistency cycle, 
fitting the hybridisation by a finite bath, and convergence of correlated properties (such as spectral functions) as a function of bath size. We should stress that similar studies were caried out before using full configuration interaction (FCI) called exact diagonalization (ED) in solid state physics community. Here, however, we will focus on using the truncated version of configuration interaction as a solver that was developed by us and examine with it the questions of interest concerning the numerics of the DMFT algorithm.  
\end{enumerate}

Our  studies are carried out on an idealised test system, namely (three-dimensional) cubic hydrogen. Hydrogen
clusters in 1, 2, and 3-dimensions have been popular models in the study of  correlation effects in quantum chemistry,
as the correlation can be tuned from the weak to the strong regime as the lattice spacing is increased \cite{johannes_dmrg_paper,scuseria_cpmft_h_cluster_paper}.  
Here we study only  cubic hydrogen (i.e. three dimensions). We use a minimal basis
(STO-3G) and a unit cell with a single hydrogen atom, and the initial Hartree-Fock crystal calculations
are carried out using the Gaussian based periodic code \textsc{Crystal}~\cite{crystal}. The use of a Gaussian basis
means that we  employ the general non-orthogonal formulation for
the Green's function quantities in section \ref{sec:summary}, as well
as the overlap corrections to the impurity model Hamiltonian and overlap in section \ref{sec:solver}. Note that 
the impurity problem in this case has only a single 1$s$ impurity orbital,
and the local Green's function also only has a single orbital index.

We begin with a brief overview of the  properties of the DMFT solution of the cubic hydrogen model before proceeding to discuss the areas above.

\subsection{The cubic hydrogen solid model}

\begin{figure*}[t]
\begin{center}
\caption{Spectral functions (density of states) from FCI, CISD, and RHF calculations for cubic hydrogen, at various lattice constants.
}
\label{fig:spec_summary}
{\bf A)} $a_{0}=1.40$ {\AA}, 9 bath orbitals, 300 frequency points. $\quad$ {\bf B)}~$a_{0}=2.25$ {\AA}, 9 bath orbitals, 300 frequency points. \\ 
{\bf C)} $a_{0}=2.50$ {\AA}, 9 bath orbitals, 300 frequency points. $\quad$ {\bf D)}~$a_{0}=6.00$ {\AA}, 9 bath orbitals, 300 frequency points. \\ 
\end{center}
\begin{tabular}{cc}
 {\bf A)} \includegraphics[scale=0.48,angle=-90,clip=true]{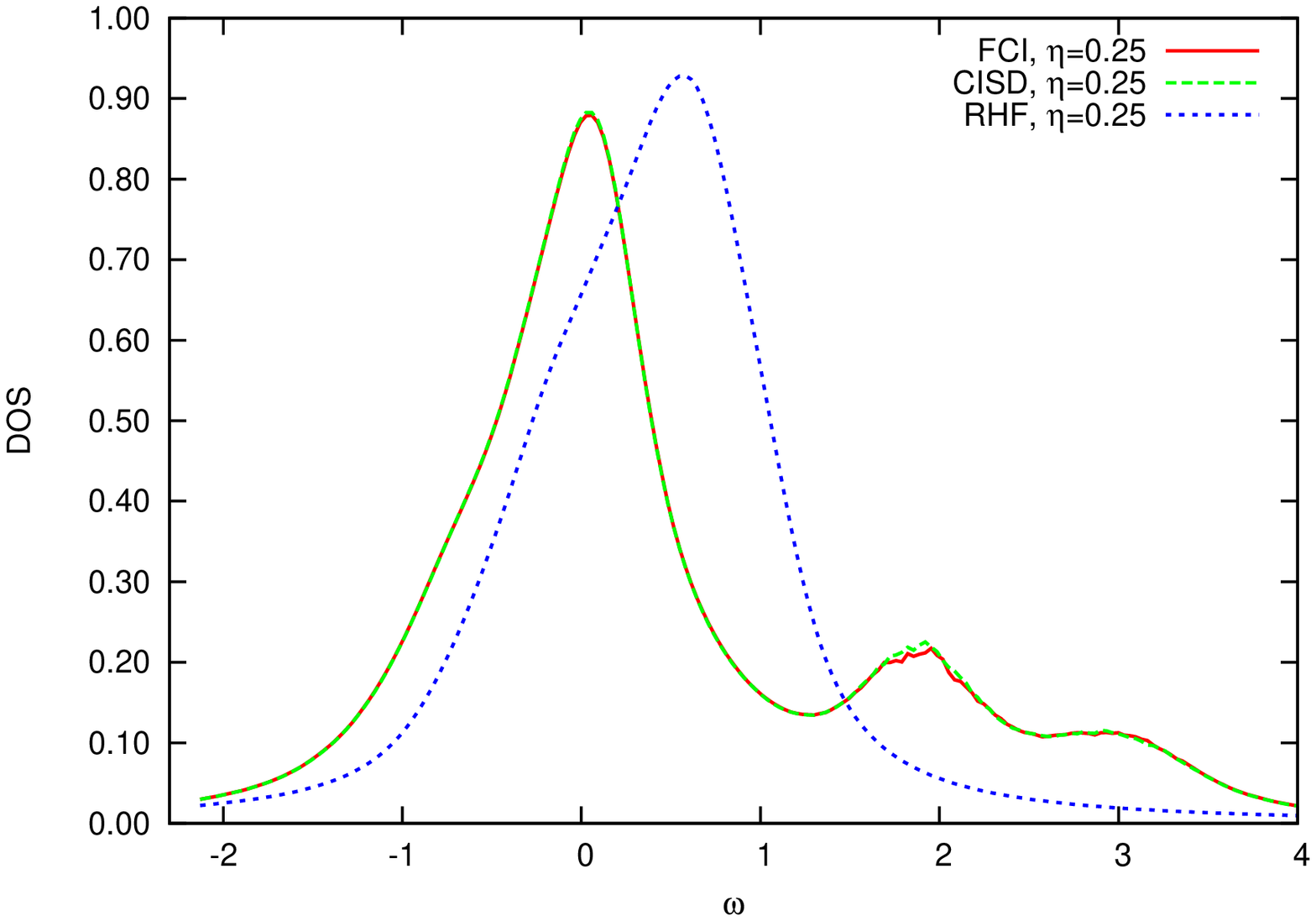} &
 {\bf B)} \includegraphics[scale=0.48,angle=-90,clip=true]{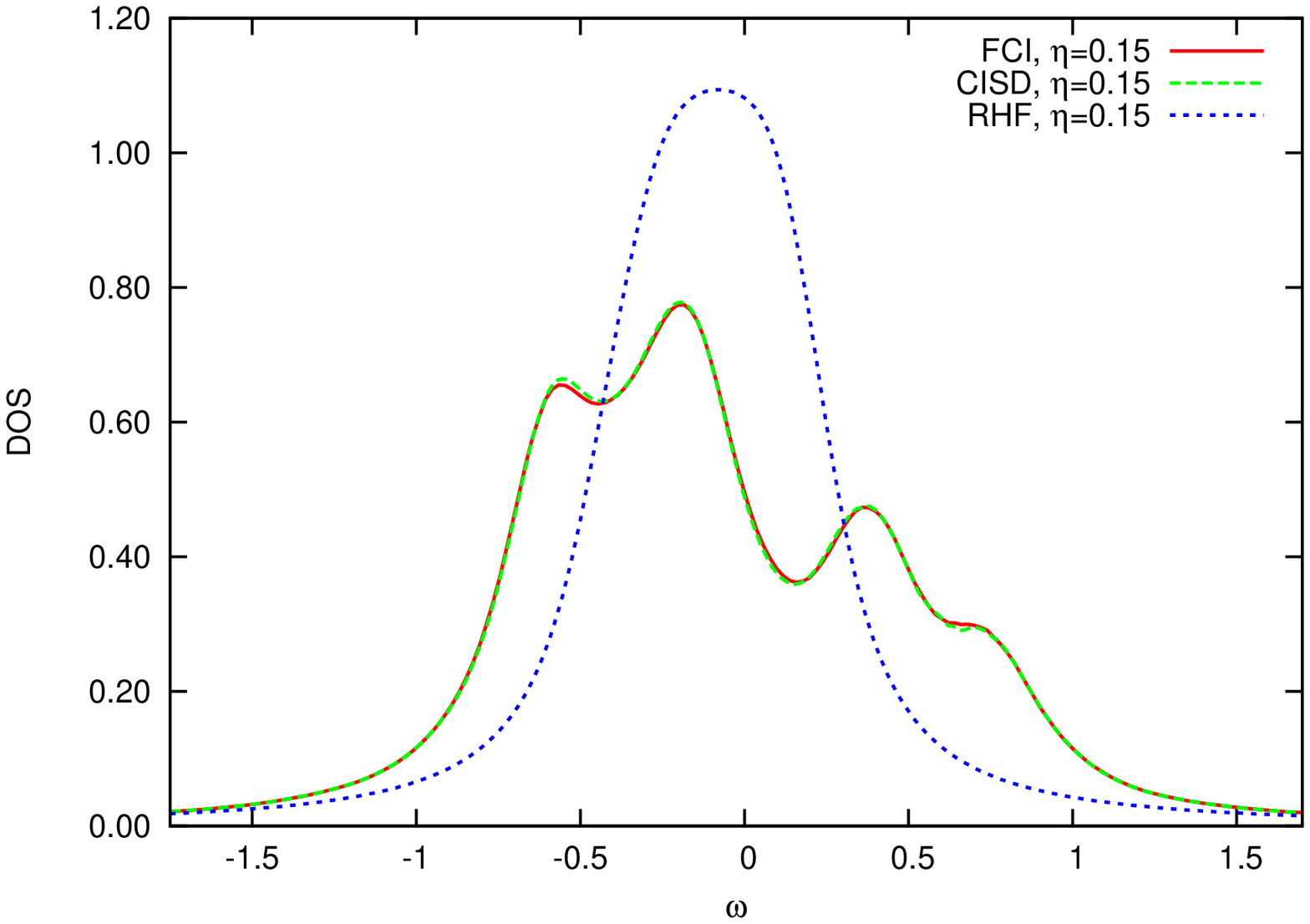} \\
 {\bf C)} \includegraphics[scale=0.48,angle=-90,clip=true]{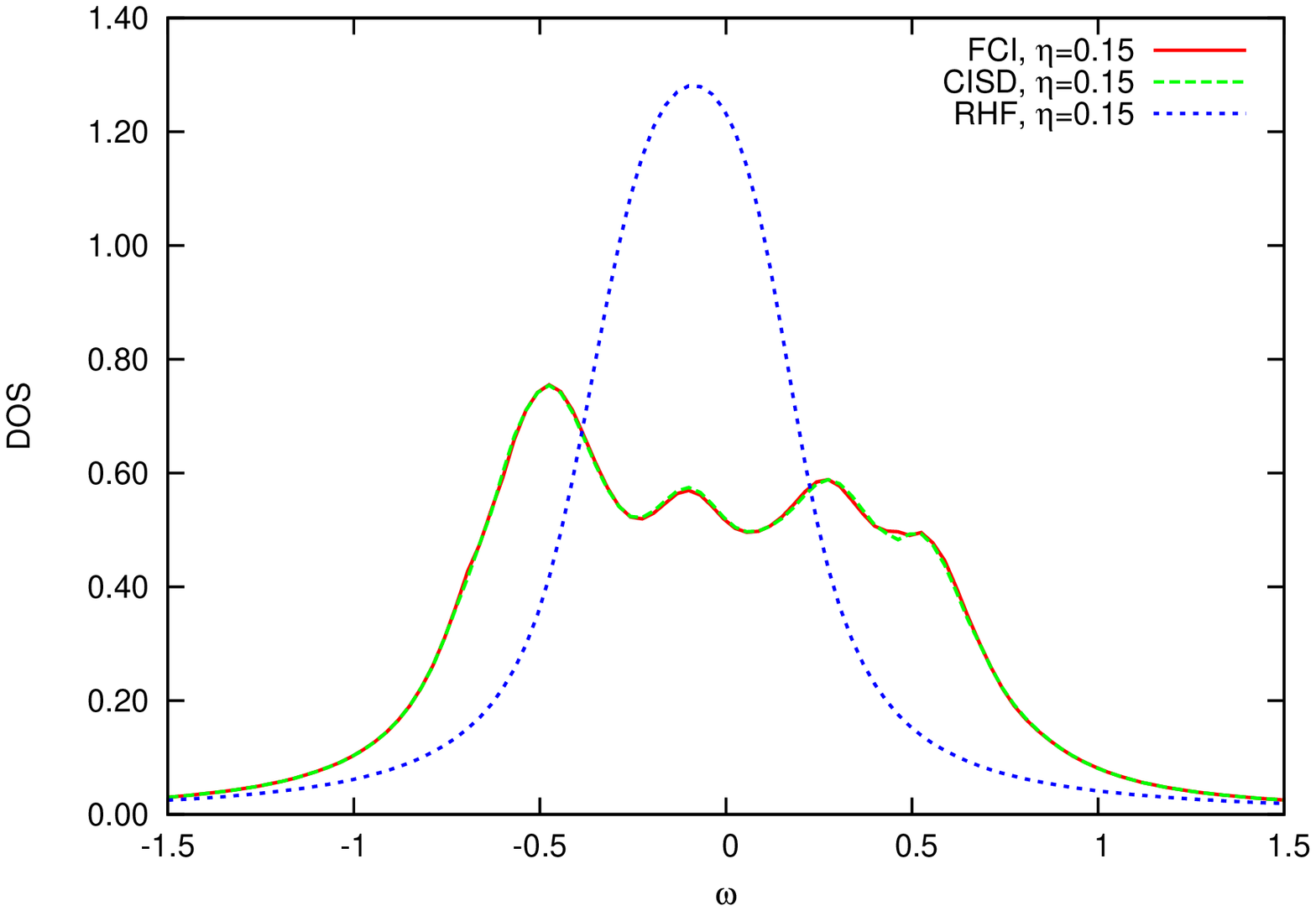} &
 {\bf D)} \includegraphics[scale=0.48,angle=-90,clip=true]{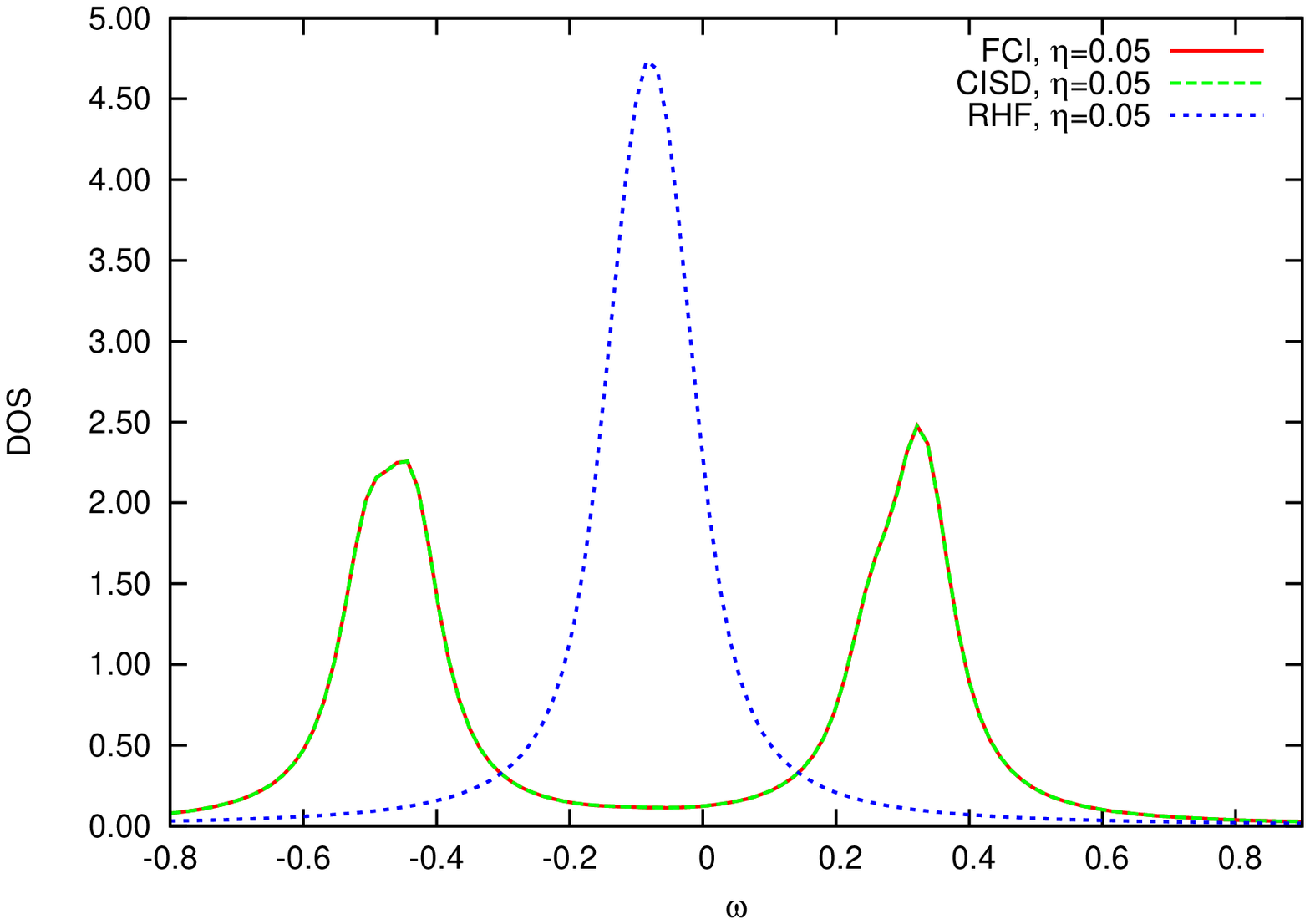} \\
\end{tabular}
\end{figure*}
\begin{table*}
\caption{Total weight of CI coefficients $c_{i}^{2}$ of different classes of determinants (Hartree-Fock (HF), singly-excited (S), 
doubly-excited (D)) in the ground-state  wavefunction of the  impurity model as a function of lattice constant $a_0$. \label{tab:det_cisd}}
\begin{tabular}{l cccc }
\hline
\hline
excitation level & $a_{0}=1.4$ & $a_{0}=2.25$ & $a_{0}=2.5$ & $a_{0}=6.0$ \\
\hline
HF        & 0.880 &  0.755 &  0.676 & 0.034\\
S         & 0.040 &  0.014 &  0.087 & 0.941\\
D         & 0.000 &  0.000 &  0.098 & 0.000\\
\hline
number of dets with $c_{i}^{2}>0.01$ & 5 & 8 & 6 & 5 \\
\hline
$\sum_{c_{i}^{2}>0.01 }c_{i}^{2}$ & 0.920 & 0.769 & 0.861 & 0.975 \\
\hline
\hline
\end{tabular}
\end{table*}

\begin{table}
\caption{Impurity model 
natural orbital occupancies for cubic hydrogen (9 bath orbitals) as a function of lattice constant $a_0$. \label{tab:nat_orb_fci}}
\begin{tabular}{l ccccccccccc }
\hline
\hline
 & level & 1-3 & 4 & 5 & 6 & 7 & 8-10 \\
\hline
$a_{0}=1.4$  & {\bf FCI}  & 2.000 &    1.999 &  1.905 &  0.095 & 0.001 &  0.000 \\
             & {\bf CISD} & 2.000 &    1.999 &  1.905 &  0.095 & 0.001 &  0.000 \\
$a_{0}=2.25$ & {\bf FCI}  & 2.000 &    1.998 &  1.718 &  0.282 & 0.002 &  0.000 \\
             & {\bf CISD} & 2.000 &    1.999 &  1.720 &  0.280 & 0.001 &  0.000  \\
$a_{0}=2.5$  & {\bf FCI}  & 2.000 &    1.999 &  1.528 &  0.472 &  0.001 &  0.000 \\
             & {\bf CISD} & 2.000 &    1.999 &  1.531 &  0.469 &  0.001 &  0.000 \\
$a_{0}=6.0$  & {\bf FCI}  & 2.000 &    2.000 &  1.000 &  1.000 & 0.000 & 0.000  \\
             & {\bf CISD} & 2.000 &    2.000 &  1.000 &  1.000 & 0.000 & 0.000  \\
\hline
\hline
\end{tabular}
\end{table}

We have carried out DMFT calculations on the cubic hydrogen model for a variety of lattice constants. 
We find that cubic hydrogen  
exhibits three  electronic regimes as a function of lattice spacing which are well-known
from analogous DMFT studies of Hubbard models \cite{RevModPhys.68.13,PhysRevLett.69.168,PhysRevLett.70.1666,PhysRevB.45.6479,Physics_Today}. We first summarise the main features of the spectral functions
and the impurity wavefunctions. (The spectral functions plotted here 
are defined as the trace of the local spectral function  in Eq. (\ref{eq:overlap_spectral})). The regimes are
\begin{itemize}
\item {\it Metallic} regime. This occurs with  lattice constants near equilibrium, and is illustrated
by  calculations at  lattice constant 1.4 {\AA}. The spectral function displays a single broad peak, indicative of
 metallic behaviour and the delocalised character of the electrons (Fig. \ref{fig:spec_summary}).
The metallic nature is also reflected in the ground-state wavefunction of the impurity model, which is  primarily a single determinant, as
seen from the natural orbital occupancies  (Table~\ref{tab:nat_orb_fci}) and from the impurity 
wavefunction determinant  analysis  (Table \ref{tab:det_cisd}).
Compared to the restricted HF spectral function, 
the correlated DMFT spectral function in Fig.~\ref{fig:spec_summary} 
displays  additional features at large frequencies and is broader, but the spectra are 
similar as expected in the weakly correlated regime. 
\item {\it Intermediate} regime. At intermediate lattice constants (e.g. 2.25 {\AA} and 2.5 {\AA})
the spectral function develops a three peak structure with features of both the metallic and insulating regime (Fig.~\ref{fig:spec_summary}).
 In early DMFT work on the Hubbard model the central peak
was a correlated feature of the spectrum not predicted in mean-field theories \cite{PhysRevLett.69.168,PhysRevB.45.6479,Physics_Today}.
The two outer peaks are  shifted from the ionisation potential and electron affinity of the atom. Analysing the impurity wavefunction,
we find that at both 2.25 {\AA} and 2.5 {\AA} lattice constants,
the wavefunction has  multideterminantal character with significant mixing of open-shell singlets and doubly excited
determinants into the ground-state (see Tables \ref{tab:det_cisd}, \ref{tab:nat_orb_fci}).  
\item  {\it Mott insulator} regime. This occurs at large lattice constants when the
hydrogen atoms  assume distinct atomic character. This is illustrated by calculations at lattice constant {6.0 \AA}. 
(In this  limit, the DMFT approximation of a local self-energy becomes exact). 
The spectral function (Fig. \ref{fig:spec_summary}) displays an insulating gap and peaks centered 
at  the electron affinity and  ionization potential of the hydrogen atom. 
The impurity wavefunction  is a mixture of open-shell singlets (see Table~\ref{tab:det_cisd}). We find that the  
singly occupied impurity natural orbitals
(Table \ref{tab:nat_orb_fci}) are respectively  localised on the impurity and the bath, thus we characterise the impurity ground-state as an impurity-bath singlet. (Note that the RHF spectral function stays metallic. An unrestricted mean-field calculation  would yield 
two peaks similar to the DMFT spectral function, but at the expense of breaking spin symmetry).
\end{itemize}

\subsection{A configuration interaction impurity solver }

\label{sec:cisolver}

As described in section \ref{sec:solver}, once the impurity model Hamiltonian has been defined, we can  determine
the impurity Green's function  within a wavefunction formalism. Here we investigate the use
of the configuration interaction (CI) hierarchy to construct impurity solvers. We can also see
this as using the DMFT framework to extend configuration interaction to the infinite system. 
To the best of our knowledge, truncated configuration interaction has not previously been explored in the DMFT literature,
although full configuration interaction (exact diagonalisation) has been widely used~\cite{RevModPhys.68.13,PhysRevB.76.245116}.
By considering CI at an  arbitrary excitation level
we obtain a hierarchy of impurity solvers that can, with increasing effort, be systematically  converged to the exact 
full CI limit, within the given bath parametrisation. 
We have based our implementation on the arbitrary excitation level CI
program in \textsc{Dalton} \cite{dalton}. Our code allows the additional possibility of defining  restricted active spaces \cite{dalton_ras}. 
However, for the simple
cubic hydrogen model,  we find that the restricted active space methodology is not necessary.
Detailed studies of the active space flexibility of the solver will thus be presented elsewhere.

To carry out CI we  define a starting determinant in a ``molecular orbital'' basis. Note that this is quite different
from how exact diagonalisation is used in DMFT, where the one-particle basis is chosen to simply be the site basis (atomic orbital basis)
of the impurity and the bath. Of course, the result of exact diagonalisation is independent of the choice of one-particle basis,
and in model problems (such as the Hubbard model), the Hamiltonian has a particularly simple local form in the site basis of the impurity and bath. However, for truncated configuration interaction the choice of starting orbital basis is of course much more important.
Here we take the molecular orbitals to be
  the eigenfunctions of the Fock operator of the impurity and bath Hamiltonian $H_{imp+bath}$, Eq. (\ref{eq:impurity_model_h}) (this
is obtained by replacing the impurity part of the Hamiltonian  by the impurity Fock operator $\mathbf{f}$ appearing in Eq. (\ref{eq:himp_gen})).
From the lowest energy orbitals we then populate a ground-state determinant and define the set of singles, doubles, and higher excited determinant spaces as in a conventional CI approximation. 
We  calculate the ground-state impurity wavefunction  
within the given CI space, generating
a CI vector $\mathbf{\psi}$ and a ground-state energy $E_{imp+bath}$. We then evaluate the  Green's function (\ref{eq:greens_def2})
 by solving the two intermediate linear equations for $\mathbf{\bar{X}}_i$, and $\mathbf{{X}}_j$
\begin{align}
&[(\omega + \mu + E_{imp+bath})\mathbf{1} - \mathbf{H}_{imp+bath})]\mathbf{\bar{X}}_i(\omega)=\mathbf{\bar{B}}_i, \\
 &\mathbf{\bar{B}}_i=\mathbf{\bar{C}}_{i} \mathbf{\psi},   \nonumber \\
&[(\omega + \mu - E_{imp+bath})\mathbf{1} + \mathbf{H}_{imp+bath})]\mathbf{{X}}_j(\omega)=\mathbf{{B}}_j, \\
&\mathbf{{B}}_j=\mathbf{{C}}_{j} \mathbf{\psi} \nonumber \label{eq:lin_eq}  
\end{align}
where $\mathbf{C}_i, \mathbf{\bar{C}_i}, \mathbf{H}_{imp+bath}$ are  representations of the impurity orbital creation, annihilation operators and 
impurity and bath Hamiltonian operator in the 
truncated CI space. (Note, for the $N+1$ and $N-1$ particle spaces accessed by the creation and annihilation operators, we consider the space of all determinants that are connected to the $N$ particle truncated CI space for the ground-state calculation). 
$\omega$ can be either purely imaginary (as used in the DMFT self-consistency cycle)
or it can be real, with a small imaginary broadening $i\eta$, when calculating the spectral function.
The Green's function matrix element is then obtained via
\begin{align}
G_{ij} = \mathbf{B}_i \mathbf{\bar{X}}_j + \mathbf{\bar{B}}_j \mathbf{X}_i
\end{align} 
The solution of the linear equations (\ref{eq:lin_eq})   can be achieved via a variety of iterative algorithms. Our implementation
follows the algorithm for CI response properties described in Ref.~\cite{ci_response}, adapted to truncated CI spaces. 

Our calculations have demonstrated that in the molecular orbital basis the modest variant of truncated configuration interaction, namly CISD, where the Hilbert space is truncated to contain only singly and doubly excited determinants, was completely sufficient to illustrate all the regimes of the hydrogen solid.    
In Fig.~\ref{fig:spec_summary} and Table ~\ref{tab:nat_orb_fci}, we show the CISD and FCI 
local spectral function and impurity natural orbital occupations in the three electronic regimes of cubic hydrogen.
In the metallic regime, the CISD spectral function is 
 completely indistinguishable from the FCI spectral function, and the same
is true for the impurity orbital natural occupation numbers. 
In the intermediate regime, for the lattice spacings 2.25 {\AA} and 2.5 {\AA}
we expect correlation effects to be stronger. However,  the impurity natural orbital occupations show that there
are only two natural orbitals with significant partial occupancy, and thus CISD is a very
good approximation to FCI. This is reflected in both the spectral functions in Fig.~\ref{fig:spec_summary} 
where CISD and FCI agree very well, as well as in the natural orbital occupation numbers,
although CISD is not as close an approximation in this case to FCI as it is in the metallic regime.
Finally, in the Mott insulator regime,  the analysis of the occupation numbers shows again that there are only two orbitals with significant
 partial occupancies and the  FCI and CISD spectral functions and impurity natural
occupation numbers are again indistinguishable.

The near-exactness of the CISD level of impurity solver is a feature of the simplicity
of the cubic hydrogen model system but also reflects the compactness of the CI expansion 
when one is using an appropriate one-particle starting basis, in this case the molecular orbital basis rather than the site basis.
 We expect that more complex solids will pose greater challenges and require
higher levels of excitation in the configuration interaction solver, and these issues will be examined elsewhere. Nonetheless,
the good performance of the single and doubles level truncation suggests that it will be promising to 
explore systematic wavefunction hierarchies in more complex problems, which may be infeasible in
the exact diagonalisation approach.

\subsection{DMFT numerics: self-consistency}

As discussed in our overview of DMFT and our specification of our implementation in section \ref{sec:dmft_algorithm}, 
the impurity model particle number $N_{imp+bath}$,  chemical potential $\mu$,
and  hybridisation $\mathbf{\Delta}(\omega)$ and  self-energy $\mathbf{\Sigma}(\omega)$ must all  be determined self-consistently. 
The determination of the optimal impurity model particle number and chemical potential are discrete and
continuous searches over single variables which are essentially robust. In contrast,  the self-consistency condition for $\mathbf{\Delta}(\omega)$ and $\mathbf{\Sigma}(\omega)$
are multi-dimensional equations. Here we examine the  convergence of the self-consistency cycle for the 
self-energy $\mathbf{\Sigma}(\omega)$ in the loop
given by steps 3.-6. in algorithm \ref{algo:two}. 

\begin{figure*}
\begin{center}
\caption{Spectral function (density of states) obtained with CISD as a solver during the iterations of the self-consistency cycle for cubic hydrogen, at
various lattice constants.\\
{\bf A)} $a_{0}=1.40$ {\AA}, 9 bath orbitals, 300 frequency points $\quad$ {\bf B)} $a_{0}=2.25$ {\AA}, 9 bath orbitals, 300 frequency points\\
{\bf C)} $a_{0}=2.50$ {\AA}, 9 bath orbitals, 300 frequency points $\quad$ {\bf D)} $a_{0}=6.00$ {\AA}, 9 bath orbitals, 300 frequency points\\
}
\label{fig:spec_sc}
\end{center}
\begin{tabular}{cc}
 {\bf A)} \includegraphics[scale=0.48,angle=-90,clip=true]{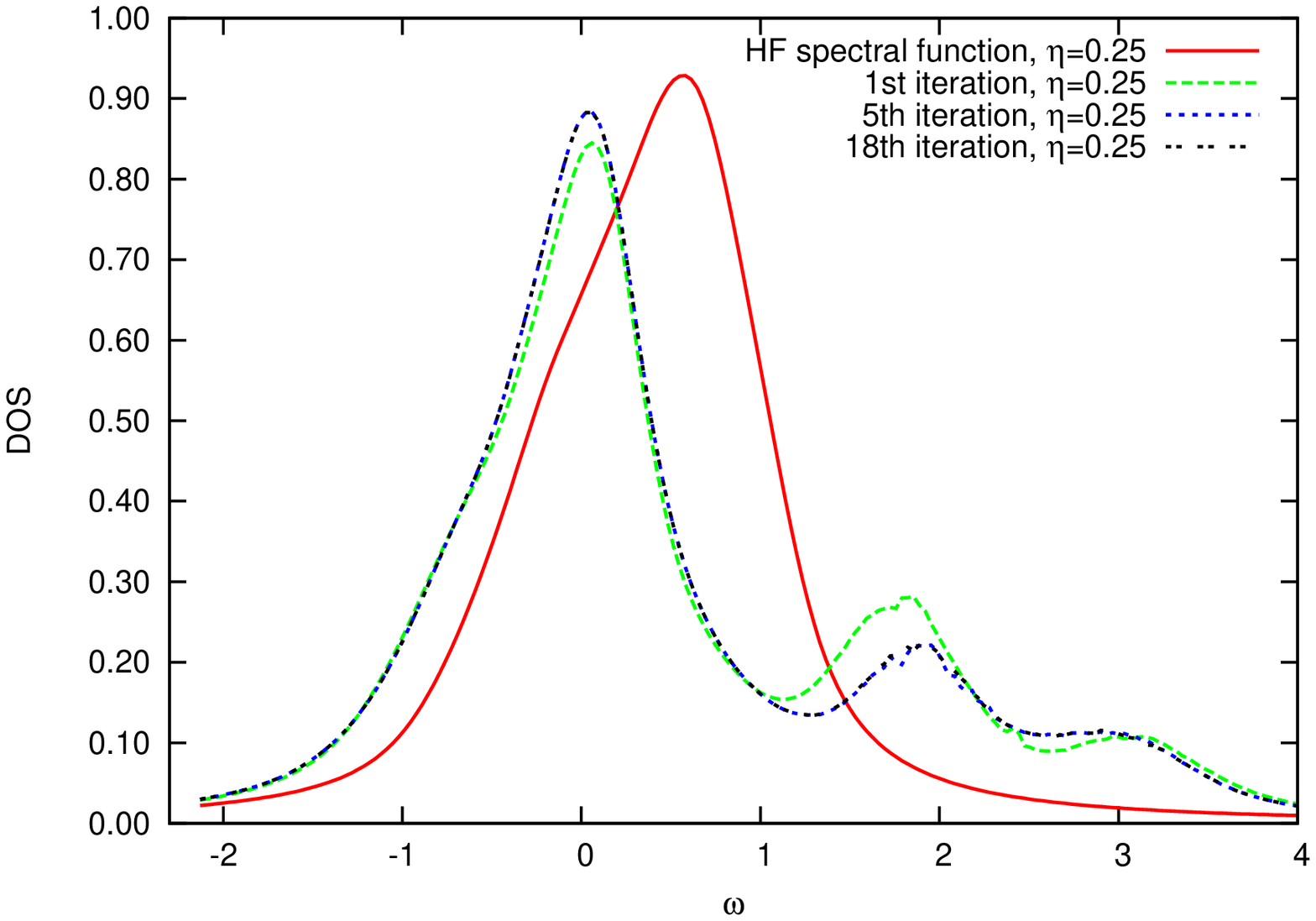} &
 {\bf B)} \includegraphics[scale=0.48,angle=-90,clip=true]{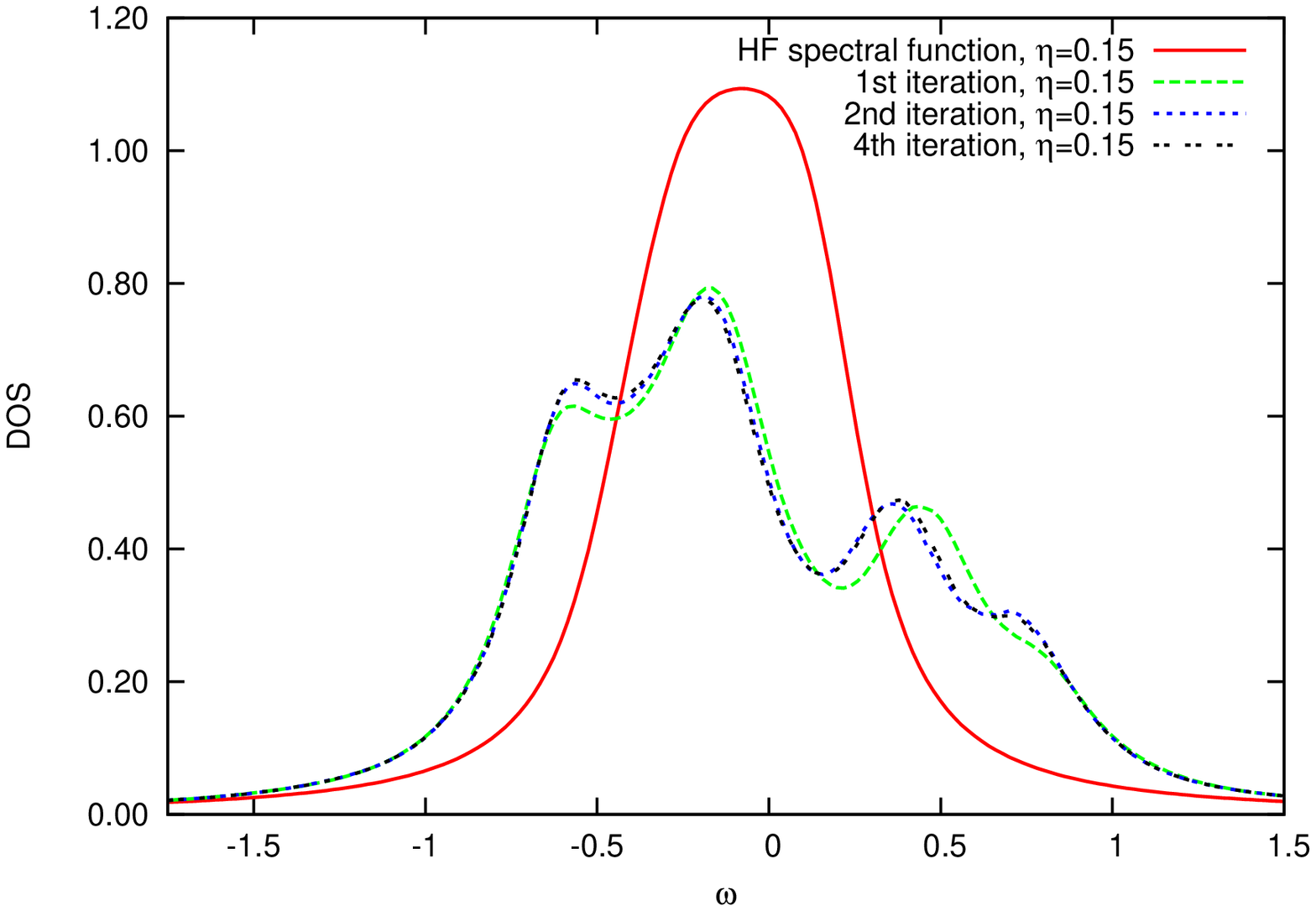} \\
 {\bf C)} \includegraphics[scale=0.48,angle=-90,clip=true]{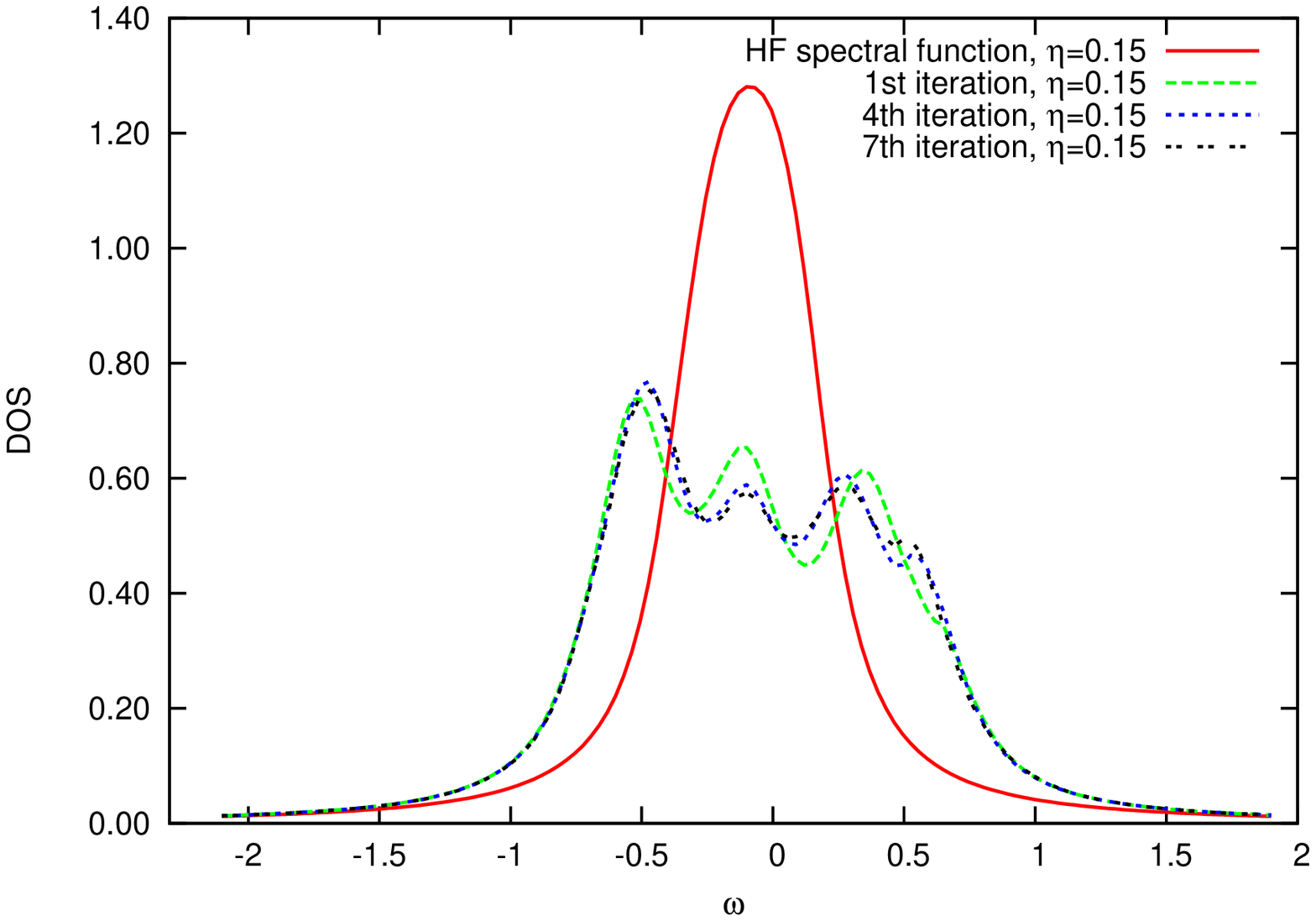} &
 {\bf D)} \includegraphics[scale=0.48,angle=-90,clip=true]{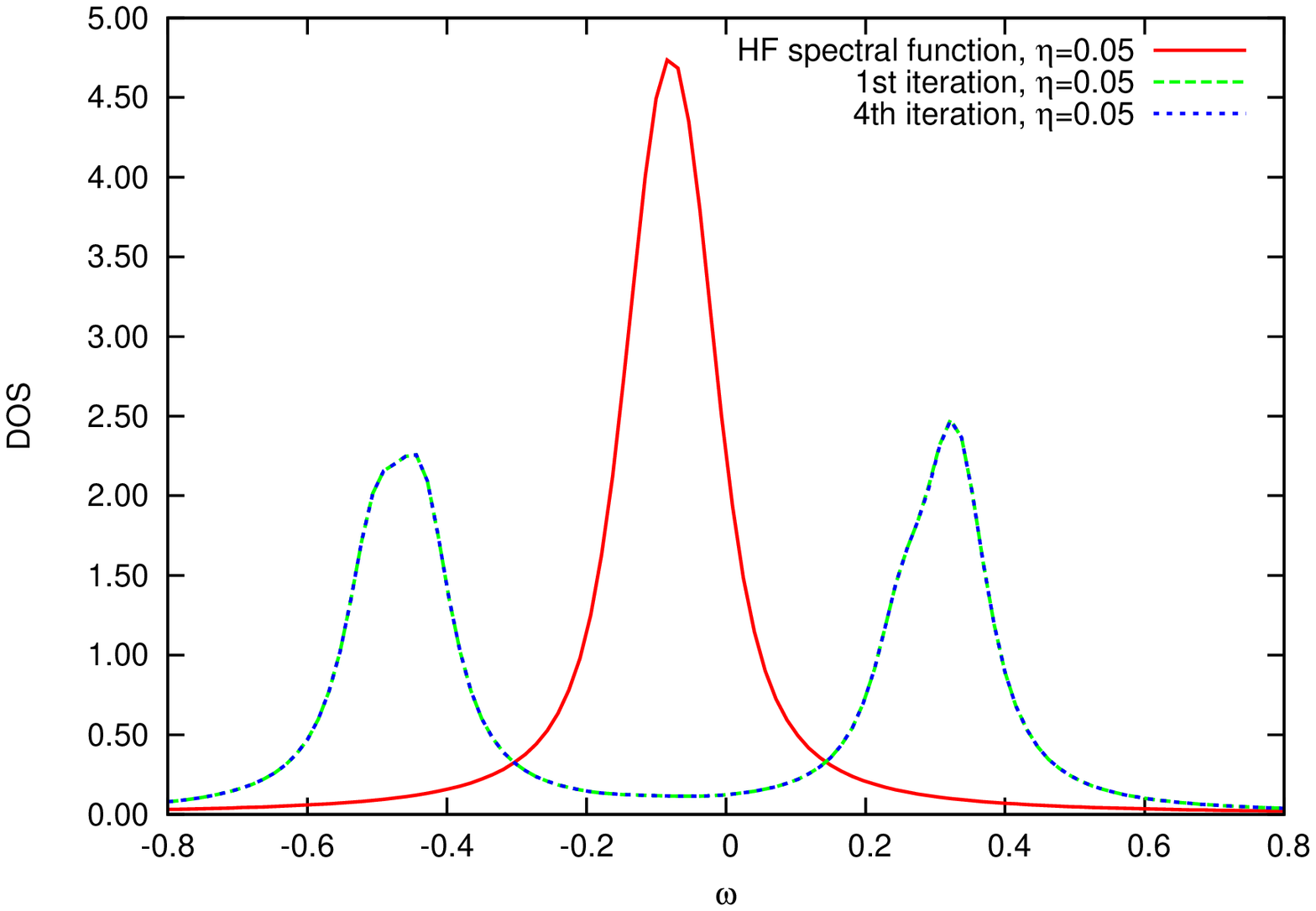} \\
\end{tabular}
\end{figure*}

In Fig. \ref{fig:spec_sc}.
we  examine the spectral functions obtained at the CISD level in the three electronic regimes of cubic hydrogen
as a function of the number of iterations of the self-consistency cycle.
Generally, we find that convergence is very rapid. In the case of the metallic regime, the spectral
function appears to converge after 5 iterations.
In the intermediate regime, for  lattice constant $a_{0}=2.25$ {\AA}

the spectral function also converges after 2 iterations. At the slightly
larger lattice constant $a_{0}=2.5$ {\AA}, convergence is a little slower and the spectral function requires 4 iterations
to converge. Finally, as we  enter the Mott insulating regime, convergence
is once again rapid and the spectral function converges after 2 iterations.

The same convergence behaviour is observed in the electronic structure of the impurity problem. In Table \ref{tab:nat_orb_kond_2.5}
we show the natural orbital occupation numbers of the impurity problem corresponding to $a_{0}=2.5$ {\AA}. These numbers were obtained using the CISD solver. (Additional tables
corresponding to the other lattice constants are given in the supplementary material~\cite{supplement1}). We see that convergence in the 2nd decimal
place is reached after 5 iterations. 
 
Overall, we  find that at least for the spectral functions of the cubic hydrogen model, only a  few iterations of self-consistency 
are  already sufficient. For quantitative properties, such as total energy evaluation of the total energy with chemical accuracy, 
we expect, however, to need a tighter convergence.

\subsection{DMFT numerics: convergence with bath size}

\label{sec:fitting}

\begin{table}
\caption{Natural orbital occupancies obtained with CISD solver during the iterations of self-consistent cycle for cubic hydrogen, $a_{0}=2.5$ {\AA}, 
9 bath orbitals, for exact parameters used to converge self-consistency see supplementary material~\cite{supplement2}.
\label{tab:nat_orb_kond_2.5}}
\begin{tabular}{l cccccc }
\hline
\hline
iter/orb no. & 1-3 & 4 & 5 & 6 & 7 & 8-10 \\
\hline
1 &  2.000  &  1.999 &  1.720 &  0.280 &  0.001 &  0.000  \\
2 &  2.000  &  1.998 &  1.583 &  0.417 &  0.002 &  0.000  \\
3 &  2.000  &  1.999 &  1.556 &  0.444 &  0.001 &  0.000  \\
4 &  2.000  &  1.999 &  1.543 &  0.457 &  0.001 &  0.000  \\
5 &  2.000  &  1.999 &  1.537 &  0.463 &  0.001 &  0.000  \\
6 &  2.000  &  1.999 &  1.533 &  0.467 &  0.001 &  0.000  \\
7 &  2.000  &  1.999 &  1.531 &  0.469 &  0.001 &  0.000  \\
\hline
\hline
\end{tabular}
\end{table}

As discussed in section \ref{sec:solver},  when dealing with an explicit bath the hybridisation $\mathbf{\Delta}(\omega)$ is 
parametrised by a finite bath, and all quantities must then be converged with respect to
the number of bath orbitals. There are  two aspects of bath convergence to explore. How difficult is the numerical problem
of fitting the hybridization to the bath couplings $\epsilon_p$ and $V_{pi}$? How rapidly
do the relevant correlated quantities (such as the DMFT spectral functions) converge with bath size? In the
latter case, the ability of the truncated configuration interaction solver (here CISD) introduced in section \ref{sec:cisolver}
to access larger bath sizes than available to exact diagonalisation, provides a new capability to examine bath convergence.

We first  discuss the numerical fitting and quality of representation  of the hybridization $\mathbf{\Delta}(\omega)$
as a function of the number of bath orbitals with couplings $\epsilon_p$ and $V_{pi}$.
We determine the  bath parameters $\epsilon_{p}$ and $V_{p i}$ by fitting $\mathbf{\Delta}(\omega)$ 
to the form (\ref{eq:bath_representation}).
 In principle, one could carry out the fit using any set of frequencies, but following standard practice,
we fit along the imaginary frequency axis, where the hybridisation is a smooth function, and use
 an equally spaced set of frequencies $\omega_n$ (Matsubara frequencies) 
\begin{equation}
\omega_{n}=\frac{(2n+1)\pi}{\beta}, \  n=0, 1, 2 \ldots
\end{equation}
where $\beta$, the inverse temperature, determines the spacing. The choice of $\beta$ is somewhat arbitrary, but to reproduce spectral
functions over a given range of frequencies, we find that it is reasonable
 to take $\beta$ to  correspond to a similar range of frequencies on the imaginary axis. 

Fitting to Eq. (\ref{eq:bath_representation}) is a highly nonlinear fit.
We find that the final fit quality
  depends  strongly on the initial choice of the parameters. We have established an initialisation procedure
to obtain a reasonable set of starting $\epsilon_p$ and $V_{pi}$,  described in the appendix. 
From this initial set, we use a Levenberg-Marquadt algorithm to minimise the metric $\sum_{nij} |\Delta_{ij}(\omega_n) - \Delta^{fit}_{ij}(\omega_n)|$
to refine the bath parameters. As described in section \ref{sec:solver}, the non-orthogonal orbital corrections for the impurity
overlap and Hamiltonian (\ref{eq:savrasov_overlap}), (\ref{eq:savrasov_hamiltonian}), are essential for obtaining a reasonable fit when the underlying crystal basis is non-orthogonal. However, we find also that if we artificially
set the  overlap matrix $\mathbf{S}(\mathbf{k})$ to the unit matrix, and proceed to fit the hybridization functions obtained in this way, considerably
 better fits are easily obtained. This suggests that it will be more efficient in the future to work within a local orthogonal basis
for the crystal, rather than the Gaussian basis currently used.

\begin{figure}
\caption{Fitting accuracy for the real part of the hybridization $Re(\mathbf{\Delta}(i\omega))$ for various numbers of bath orbitals. The number of frequencies employed was 128 and $\beta=128$.  \label{fig:bath_1site_re}}
\includegraphics[angle=-90,width=0.95\columnwidth]{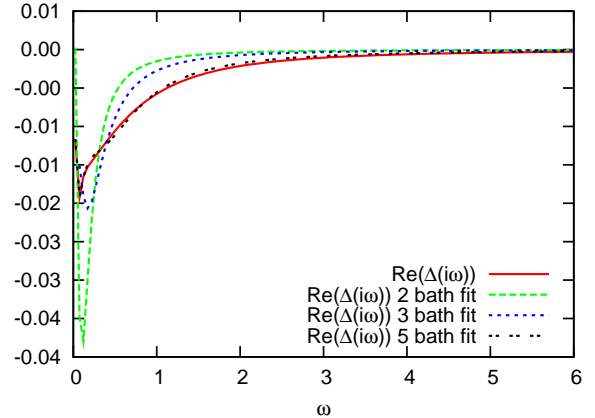}
\end{figure}

\begin{figure}
\caption{Fitting accuracy for the imaginary part of the hybridization $Im(\mathbf{\Delta}(i\omega))$ for various numbers of bath orbitals. The number of frequencies employed was 128 and $\beta=128$.  \label{fig:bath_1site_im}}
\includegraphics[angle=-90,width=0.95\columnwidth]{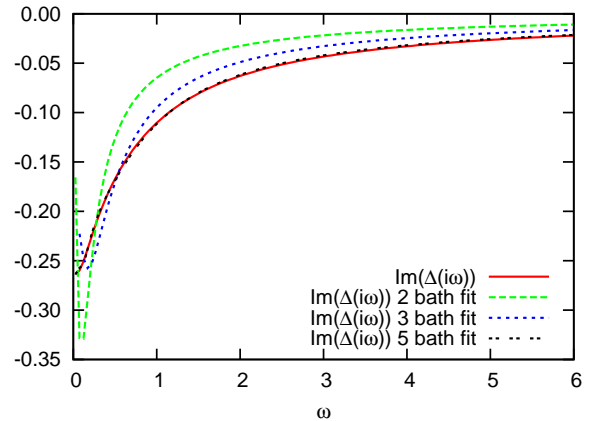}
\end{figure}

\begin{figure}
\caption{Fitting accuracy with different number of bath orbitals for the Hartree-Fock impurity spectral function of cubic hydrogen.
The number of frequencies employed was 128 and $\beta=128$.  \label{fig:bath_1site_spectral}}
\includegraphics[angle=-90,width=0.95\columnwidth]{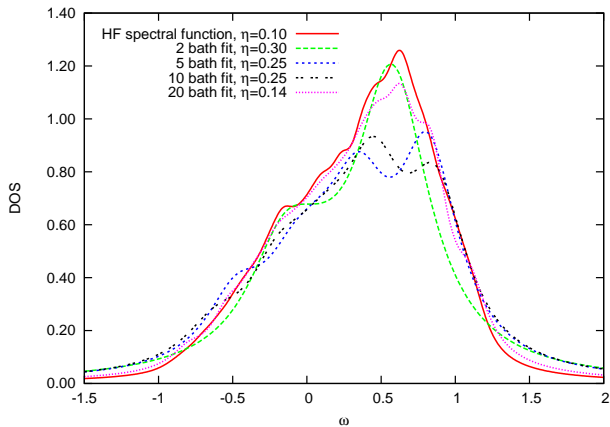}
\end{figure}

We show the results of the fitting procedure for the real and imaginary parts
of the Hartree-Fock hybridization (defined in section \ref{sec:dmft_algorithm}) in  the metallic regime  in Fig.~\ref{fig:bath_1site_re} 
and Fig.~\ref{fig:bath_1site_im}. Similar studies of illustrating difference between Green's functions obtained for different number of bath orbitals can be found in Appendix~C of Ref.~\cite{RevModPhys.68.13} or for cluster DMFT in Ref.~\cite{PhysRevB.78.115102}.
It is evident that the fit  becomes better as we increase the number of bath orbitals, and indeed with 5 bath orbitals the fits appear
exact to the eye. However, the quality 
of the fit along the imaginary axis does not necessarily guarantee the same quality of reproduction of properties along the real axis.
In Fig. ~\ref{fig:bath_1site_spectral} we show the 
 convergence of the accuracy of the impurity spectral function, $-\frac{1}{\pi} \Im \mathrm{Tr} \mathbf{G}_{imp}(\omega)$
to the corresponding Hartree-Fock quantity $-\frac{1}{\pi} \Im \mathrm{Tr} \mathbf{g}(\mathbf{R}_0, \omega)$. (Note that
this is not the physical local spectral function, which must be defined in a non-orthogonal basis which an additional overlap factor, as in Eq. (\ref{eq:overlap_spectral})).
For two bath orbitals the fit on the imaginary axis is poor  and  the spectral function on 
the real axis is  poorly represented as well. Once the number of bath orbitals is increased to five orbitals, the error of the fit on the imaginary 
axis becomes quite small and the spectral function becomes appropriately improved. 
However, the rate of the improvement of the spectral function with respect to the number of bath
orbitals is  slower than the improvement of the fit on the imaginary axis, as it is much less smooth. 
Note that for each of the spectral functions in Fig.~\ref{fig:bath_1site_spectral} we have chosen a different broadening parameter $\eta$
to reflect the changing bath orbital spacing.

\begin{figure}
\caption{Spectral function (density of states) obtained with CISD solver for different number of bath orbitals for cubic hydrogen, $a_{0}=2.25$ {\AA}. \label{fig:bath_conv_kond_2.25}}
\includegraphics[angle=-90,width=0.95\columnwidth]{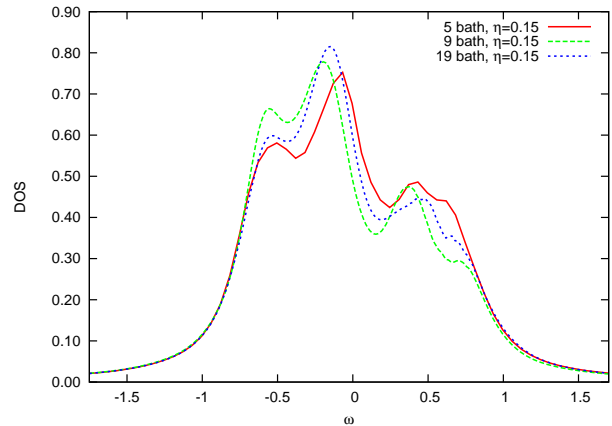}
\end{figure}

We now turn to the convergence of the correlated DMFT quantities as a function of bath size. The need
to examine this convergence is an essential feature of working within the discrete bath formulation.
In Fig. \ref{fig:bath_conv_kond_2.25}  we present
the cubic hydrogen local spectral functions obtained using the CISD method as a solver at lattice constant 2.25 {\AA}  using 
5, 9, and 19 bath orbitals in the impurity model, the latter bath size being comfortably beyond what
can be studied using exact diagonalisation. In addition, in Table \ref{tab:nat_orb_bath} we also present 
the impurity natural occupation numbers calculated with CISD solver
with the different bath sizes as a more quantitative test of the bath size convergence. Similar studies of the convergence of the occupation numbers with respect of to the bath size while using exact diagonalization as a solver can be found in Ref.~\cite{PhysRevB.69.195105,PhysRevB.78.115102}.

We see that the spectral functions are in fact quite similar between the different bath sizes. Indeed
already the very small 5 bath orbital result is remarkably similar to the 19 bath orbital result. This
must be considered  a feature of the simplicity of the cubic hydrogen model which has only a single orbital
in the unit cell. Examining the impurity model 
 natural occupation numbers we also see that all bath orbital sizes yield very similar natural occupancies with
only very small differences.
This is promising for future applications as it seems only a relatively small number of bath orbitals is necessary
to obtain a converged result.
\begin{table}
\caption{Impurity natural orbital occupancies obtained with CISD solver for cubic hydrogen at 
lattice constants 2.25 {\AA}, using 5, 9, and 19 bath orbitals. \label{tab:nat_orb_bath}}
\begin{tabular}{l| ccccccccc }
\hline
\hline
5 bath      &    1   &    2   &    3   &      4   &      5  &    6   \\
$a_{0}=2.25$&  2.000  & 1.999 &  1.710 &    0.290 &   0.001 &   0.000 \\
\hline
9 bath        &  1-3  &    4   &    5   &    6   &   7   &    8-10  \\
$a_{0}=2.25$  & 2.000 &  1.999 & 1.720  & 0.280 &  0.001 &   0.000\\   
\hline
19 bath      &  1-8  &    9   &  10   &   11  &   12   &    13-20 \\
$a_{0}=2.25$ & 2.000 &  1.999 &  1.739 & 0.261 & 0.001 &   0.000 \\
\hline
\hline
\end{tabular}
\end{table}

\section{Conclusions}

\label{sec:conclusions}

In this work we have carried out an initial study of dynamical mean-field theory (DMFT) 
from a quantum chemical perspective.
DMFT provides a powerful framework  to extend  quantum chemical correlation hierarchies
to infinite problems through a self-consistent embedding view of the crystal. The
basic approximation is one of a local self-energy, which is a kind of 
local correlation approximation.

We have explored several ways in which quantum chemical ideas 
can be combined with the DMFT framework. First, we start
with a Hartree-Fock based DMFT Hamiltonian which avoids the double counting problems of the 
 commonly employed DFT-DMFT scheme. 
Second, we have investigated the truncated configuration interaction (CISD) as an impurity solver.
 The CI hierarchy avoids the sign problem
 inherent to  Monte Carlo solvers in DMFT, and allows a systematically improvable 
approach to the exact solution. Conversely,
the DMFT framework enables even truncated CI to be extended to the infinite crystal. 
In the simple but challenging
cubic hydrogen model we find that CI at the singles and doubles level  already 
reproduces the 
 structure of the density of states in the various electronic regimes with near perfect accuracy.
 Finally, we have carried out an investigation of some  numerical aspects of the DMFT procedure,
including convergence of the self-consistent cycle and convergence of properties with respect to the bath
discretisation.  We find that modest bath sizes, easily accessible to the CI solver, already produce converged results.

These investigations should be viewed as first steps and there are many avenues to develop these ideas.
For example, the  Hartree-Fock starting point in DMFT treats long-range Coulomb interactions at only
 the mean-field level, neglecting  long-range screening. Quantum chemical perturbation techniques
may be useful in treating these additional interactions and may prove complementary to current Green's function treatments
of screening \cite{crystal_mp2,aryasetiawan_gw_dmft}. Also, there is a wealth of quantum chemical wavefunction approximations that could be combined with
the DMFT framework, the most obvious example being coupled cluster theory, which should prove
advantageous over configuration interaction as the number of impurity orbitals increases. 

Additionally, the main ideas in this work, in particular, the use of
quantum chemical Hamiltonians
and solvers, are not limited to the single orbital DMFT that we have
used to study cubic hydrogen. Their combination with multi-orbital and
cluster versions of DMFT~\cite{RevModPhys.77.1027,PhysRevLett.87.186401,PhysRevB.61.12739,PhysRevB.58.R7475} should be investigated. 
Finally, the
possibility of using DMFT in finite systems,
either within the standard DMFT formalism~\cite{PhysRevLett.103.016803,PhysRevB.82.195115}  or through
a true finite DMFT formalism~\cite{millis_preprint}, or
the use of DMFT ideas with quantum variables other than the Green’s
function are further intriguing possibilities for the future.

\section{Acknowledgments}

This work was supported by the  Department of Energy, Office of Science. We acknowledge useful conversations with A. J. Millis, C. A. Marienetti, D. R. Reichman, and G. Kotliar.

\section{Appendix: Guess for bath fitting}

To generate some initial guess bath parameters $\epsilon_p$ and $V_{pi}$ for the bath fitting, we follow the procedure below. Let
us specialise to the case of a single impurity orbital where we can drop the $i$ index. Then
the bath parametrisation (\ref{eq:bath_representation}) becomes
\begin{align}
\Delta(\omega_n)=\sum_p \frac{V_p^2}{\omega_n - \epsilon_p}
\end{align}
where we have assumed $V_p$ is real. Viewing $1/(\omega_n - \epsilon_p)$ as the elements of a matrix $M_{np} = 1/(\omega_n - \epsilon_p)$,
the above becomes the matrix equation
\begin{align}
\Delta_n = \sum_p M_{np} W_p
\end{align}
where $\Delta_n=\Delta(\omega_n)$ and $W_p = V_p^2$. We can invert this equation to obtain the couplings
\begin{align}
W_p = \sum_n M^{-1}_{pn} \Delta_n
\end{align}
where we understand $\mathbf{M}^{-1}$ to mean the generalised inverse in the singular value decomposition sense.
There are now only two remaining issues. First, we have to choose a set of $\epsilon_p$ to define the matrix $\mathbf{M}$. Second, 
given arbitrary $\Delta_n$, $W_p$ is not necessarily positive definite (and thus does not necessarily yield real couplings $V_p$). We
find the latter to be a problem particularly when the overlap matrix (due to non-orthogonality) is significantly
different from unity, which further suggests (as discussed in section \ref{sec:fitting}) that it will be advantageous to work in an orthogonal basis in the future.

In the first case, we take roots of the Legendre polynomial of order $P/2$ where $P$ is the number of bath levels we wish to fit
and map them respectively from the $[-1,1]$ interval (associated with the Legendre roots) to $[0, \infty]$ and $[-\infty, 0]$ using
the transformation ${1-x}/(\lambda(1+x))$, where $\lambda$ is a scaling factor that is optimised to produce the best fit.
In the second case, we simply take $V_p=\Re({W_p^{-1/2}})$.

\bibliographystyle{/usr/share/texmf/bibtex/bst/revtex/apsrev.bst}
\bibliography{dmft}

\newpage

\section{Supplementary material}

\subsection{Impurity natural orbital occupancies obtained with CISD solver during the iterations of self-consistent cycle for cubic hydrogen at various lattice constants,  9 bath orbitals.}

\begin{table}[h!]
\caption{$a_{0}=1.4$ $\mathring{A}$}
\begin{tabular}{l cccccccc} 
\hline
\hline
iter/orb no. & 1-3 & 4 & 5 & 6 & 7 & 8 & 9-10 \\
\hline
1    &  2.000  &  1.999 &  1.908 &  0.091 &  0.001 &  0.001  &  0.000\\
2    &  2.000  &  1.999 &  1.886 &  0.113 &  0.001 &  0.001  &  0.000\\
3    &  2.000  &  1.999 &  1.906 &  0.094 &  0.001 &  0.000  &  0.000\\
4    &  2.000  &  1.999 &  1.902 &  0.098 &  0.001 &  0.000  &  0.000\\
5    &  2.000  &  1.999 &  1.898 &  0.101 &  0.001 &  0.001  &  0.000\\
6    &  2.000  &  1.999 &  1.905 &  0.094 &  0.001 &  0.001  &  0.000\\
7    &  2.000  &  1.999 &  1.905 &  0.094 &  0.001 &  0.001  &  0.000\\
8-20 &  2.000  &  1.999 &  1.905 &  0.095 &  0.001 &  0.000  &  0.000\\
\hline
\hline
\end{tabular}
\caption{$a_{0}=2.25$ $\mathring{A}$}
\begin{tabular}{l ccccccc }
\hline
\hline
iter/orb no. & 1-3 & 4 & 5 & 6 & 7 & 8-10 \\
\hline
1 &    2.000 &  1.999 &  1.801 &  0.199 &  0.002 &   0.000 \\
2 &    2.000 &  1.999 &  1.742 &  0.258 &  0.001 &   0.000 \\
3 &    2.000 &  1.999 &  1.725 &  0.275 &  0.001 &   0.000 \\
4 &    2.000 &  1.999 &  1.720 &  0.280 &  0.001 &   0.000 \\
5 &    2.000 &  1.999 &  1.720 &  0.280 &  0.001 &   0.000 \\
\hline
\hline
\end{tabular}
\caption{$a_{0}=6.0$ $\mathring{A}$}
\begin{tabular}{l ccccccc }
\hline
\hline
iter/orb no. & 1-3 & 4 & 5 & 6 & 7 & 8-10 \\
\hline
1 &    2.000 &  1.998 &  1.160 &  0.840 &  0.002 &  0.000 \\
2 &    2.000 &  2.000 &  1.000 &  1.000 &  0.000 &  0.000 \\
3 &    2.000 &  2.000 &  1.000 &  1.000 &  0.000 &  0.000 \\
4 &    2.000 &  2.000 &  1.000 &  1.000 &  0.000 &  0.000 \\
\hline
\hline
\end{tabular}
\end{table}

\subsection{Calculation details}
\begin{itemize} 
\item 5 bath dmft self-consistency using CISD solver was carried out for 200 imaginary frequencies and $\beta=100$, the used damping factor was $\alpha=0.4$, convergence threshold on self-energy $\tau=0.005$. 
\item 9 bath dmft self-consistency using both CISD and FCI solvers for all the lattice constants was carried out for 200 imaginary frequencies and $\beta=100$, the used damping factor was $\alpha=0.7$, convergence threshold on self-energy $\tau=0.005$. 
\item 19 bath dmft self-consistency using CISD solver was carried out for 200 imaginary frequencies and $\beta=100$, the used damping factor was $\alpha=0.8$, convergence threshold on self-energy $\tau=0.009$.  
\end{itemize}

\end{document}